\documentclass[twocolumn,aps,prb,showpacs,floatfix]{revtex4}

\usepackage{graphicx,amssymb,amsmath}

\newcommand{\rhomb}{\text{\protect\rotatebox[origin=c]{45}{$\square$}}}

\begin{document}

\title{The Ioffe-Regel criterion and diffusion of vibrations in random lattices}
\author{Y. M. Beltukov}
\author{V. I. Kozub}
\affiliation{A. F. Ioffe Physical-Technical Institute, 194021
Saint Petersburg, Russia}
\author{D. A. Parshin}
\affiliation{Saint Petersburg State Polytechnical University,
195251 Saint Petersburg, Russia}

\date{\today}

\begin{abstract}
We consider diffusion of vibrations in  $3d$ harmonic lattices
with strong force-constant disorder. Above some frequency
$\omega_{\rm IR}$, corresponding to the Ioffe-Regel crossover,
notion of phonons becomes ill defined. They cannot propagate
through the system and transfer energy.  Nevertheless most of the
vibrations in this range are not localized. We show that they are
similar to {\em diffusons} introduced by Allen, Feldman et al.,
Phil. Mag. B {\bf 79}, 1715 (1999) to describe heat transport in
glasses. The  crossover frequency $\omega_{\rm IR}$ is close to
the position of the boson peak. Changing strength of disorder we
can vary $\omega_{\rm IR}$ from zero value (when rigidity is zero
and there are no phonons in the lattice) up to a typical frequency
in the system.  Above $\omega_{\rm IR}$ the energy in the lattice
is transferred by means of diffusion of vibrational excitations.
We calculated the diffusivity  of the modes $D(\omega)$ using both
the direct numerical solution of Newton equations and the formula
of Edwards and Thouless. It is nearly a constant above
$\omega_{\rm IR}$ and goes to zero at the localization threshold.
We show that apart from the diffusion of energy, the diffusion of
particle displacements in the lattice takes place as well. Above
$\omega_{\rm IR}$ a displacement structure factor $S({\bf q},
\omega)$ coincides well with a structure factor of random walk on
the lattice.  As a result the vibrational line width
$\Gamma(q)=D_u q^2$ where $D_u$ is a diffusion coefficient of
particle displacements.  Our findings may have important
consequence for the interpretation of experimental data on
inelastic x-ray scattering and mechanisms of heat transfer in
glasses.
\end{abstract}

\maketitle

\section{Introduction}

Propagation of vibrational excitations in disordered systems is
one of the advanced problems in condensed matter physics. In
particular, transport mediated by these excitations is responsible
for the thermal conductivity of amorphous dielectrics (glasses).
However mechanisms of heat transfer in glasses above the plateau
region are still poorly understood.

At low temperatures below 1\,K the low frequency plane long wave
acoustical phonons are well defined excitations which transfer the
heat in glasses. At these temperatures the thermal conductivity
$\varkappa(T)\propto T^2$ and is controlled by a resonant
scattering of phonons on two-level systems (TLS)~\cite{Hunklinger,
Phillips}. Between 4\,K and 20\,K the thermal conductivity
$\varkappa(T)$ saturates and displays a well known
plateau~\cite{Zeller}. As was shown in~\cite{Buchenau1} it can be
explained by resonant scattering of phonons by quasilocal
vibrations (QLV). The QLV, together with TLS and phonons are
vibrational excitations responsible for many universal properties
of glasses~\cite{Parshin1}.

However, above approximately 20\,K the thermal conductivity rises
again and finally saturates on the level of one order of magnitude
higher, at temperatures about several hundreds
Kelvin~\cite{CahillPohl}. As generally believed, the origin of
this second rise of the thermal conductivity (above the plateau)
is not related to phonons. It was established long
ago~\cite{BirchClark, Kittel, Graebner}, that in this temperature
(frequency) range the mean free path of phonons $l$ becomes of the
order of their wave length $\lambda$ (or even smaller, of the
order of interatomic distance). Correspondingly, the Ioffe-Regel
criterion for phonons~\cite{Ioffe} becomes violated. The existence
of such crossover was confirmed by molecular dynamics calculations
for some real and model glasses~\cite{taraskin, schober} and
disordered lattices~\cite{schirm, taras}.

In the regime of such strong scattering a standard concept of
plane waves (phonons) with a well defined wave vector $\bf q$
becomes inapplicable. The question then arises: what physical
mechanism is responsible for the heat transfer in glasses in this
temperature range? The numerical simulations show that majority of
the vibrational modes in the corresponding frequency range are not
localized~\cite{jin, oligschleger, taraskin2}.

As was shown in~\cite{Cahill1, Cahill2, Cahill3}, a lower limit of
the thermal conductivity of amorphous solids above 30 K can be
correctly estimated within the framework of the Einstein's
model~\cite{Einstein}. It was assumed that the mechanism of heat
transport above the plateau is a random walk of thermal energy
between clusters of neighboring atoms vibrating with random
phases. In fact, a diffusion mechanism for the heat transfer in
this temperature range was proposed.

At the same time, delocalized vibrations in glasses of  a new
type,  different from phonons, were introduced. They were called
{\em diffusons}~\cite{Nature1, Nature2, Nature3, Nature4,
Nature5}. These are vibrations spreading through the system not
ballistically, as phonons (on distances of the order of mean free
path) but by means of diffusion. It is an important class of
excitations which  occupy in glasses the dominant part of the
spectrum~\cite{Nature5}. In these papers it was put forward the
hypothesis that the boundary between phonons and diffusons is
determined by the Ioffe-Regel criterion for phonons. Since
diffusons are delocalized excitations, they may be responsible for
the thermal conductivity of glasses above the plateau.

The similar conclusion was made by the authors of~\cite{strong1,
strong2}. They considered the case of strong scattering of phonons
in disordered lattices with a significant fraction of randomly
located missing sites, but which is still far from the percolation
threshold. It was shown that, in contrast to the electronic case,
the Ioffe-Regel criterion is inaccurate in the prediction of
phonon localization. Instead of localization, the vibrational
transport above the Ioffe-Regel threshold becomes diffusive with
approximately constant energy diffusivity $D(\omega)$. The
diffusivity was calculated by numerical solution of the Newton
equations for particle displacements. Similar calculations but for
real glasses were done in the papers~\cite{sim, similar} using
molecular dynamics methods.

The diffusons above the Ioffe-Regel crossover were identified also
in granular jammed systems with repulsive forces between the
particles~\cite{jammed1, jammed2}. They also have diffusivity
which is independent of frequency $\omega$. It was calculated
making use of the Kubo-Greenwood formula for the thermal
conductivity derived in~\cite{Nature2}. In jammed systems the
Ioffe-Regel crossover frequency $\omega_{\rm IR}$ can vary. It is
shifted to zero when the system approaches the jamming transition
point and rigidity goes to zero.

Therefore, as we believe, it is important to study properties of
diffusons systematically in systems where they exist. They bring a
new physics to our understanding of vibrational properties in
strongly disordered systems and energy/heat transfer in glasses.
To study these properties, we should have a model being
sufficiently simple but still allowing to describe all of them.

Since we consider harmonic models, the simplest but still rather
general, are (scalar or vector) models  where particles, placed in
equilibrium positions, are connected by random elastic springs.
The equilibrium positions can be taken on a lattice~\cite{schirm,
taras, bunde, taraskin3}, or randomly distributed in
space~\cite{parisi1, grigera}. With some exceptions, there is no
principal difference between these two cases because equilibrium
positions do not enter to the dynamical matrix. The only important
features are the type of disorder in elastic spring constants and
the topology of the bonds. If all spring constants are positive,
one can study different situations taking different distributions
of random springs to explain existing experimental
data~\cite{bunde}.

However a problem appears when some spring constants take negative
values~\cite{schirm, taraskin3, grigera}. Even if the number of
negative springs and their absolute values are relatively small,
in the absence of any correlation with positive springs they
produce an inevitable mechanical instability in all infinite
systems. Therefore, a situation of strong force-constant disorder
when appreciable concentration of springs can take negative values
comparable with positive ones, is not possible in these models. On
the other hand, an inclusion of negative spring constants can
considerably improve an agreement between theory and existing
experimental data~\cite{schirm, taraskin3, grigera}.

One can solve this problem mathematically, using a stable random
matrix approach~\cite{PositiveDefine,Chalker,our1,our2}. In this
approach positive and negative springs are tangly correlated with
each other. These correlations automatically guaranty the
mechanical stability of the system independently of character and
strength of the force constant disorder. In the present paper we
are going to use this approach to investigate diffusion of
vibrational excitations in disordered lattices with strong
force-constant disorder. Some of our preliminary results have been
presented in short form elsewhere~\cite{our2}.

The paper is organized as follows. In  Section~\ref{rma} for the
sake of clarity of further consideration, we outline the main
properties of the model. We consider disordered lattices with
strong force-constant disorder, described by stable positive
definite random dynamical matrix $AA^T$ having positive
eigenvalues only. Matrix $A$ is a random matrix (not necessary
symmetric) built on simple cubic lattice, with statistically
independent matrix elements between the nearest neighbors $A_{i\ne
j}$, having zero mean $\left<A_{i\ne j}\right>=0$ and equal
variance $\left<A_{i\ne j}^2\right>=V^2$.   We show that the
density of states $g(\omega)$ is not zero at $\omega=0$ and
phonons cannot propagate through the lattice. Similarly to systems
at jamming transition point, the rigidity  of the lattice is also
zero. However the physical reason is different. In our case it is
due to  high concentration of negative springs (about 45\%) in the
system what makes it extremely soft. The participation ratio
$P(\omega)$ indicates that all modes with exception of high
frequency part are delocalized. As it is shown by further
investigation, all of them are diffusons. In Section~\ref{phonons}
we consider slightly additively deformed dynamical matrix
$AA^T+\mu M_0$ which has phonon-like excitations at small
frequencies. Here positive definite matrix $M_0$ (random or
non-random) is independent of $A$. $\mu$ is a parameter of the
model which can vary in the interval $0\leqslant\mu < \infty$.
Analyzing properties of this matrix, we calculate the Young
modulus $E$, sound velocity, the density of states and
participation ratio, the dynamical structure factor $S({\bf q},
\omega)$, the phonon dispersion law $\omega_{\bf q}$, and also
their mean free path $l(\omega)$. Comparison of the later with
phonon wave length $\lambda$ determines the Ioffe-Regel crossover
frequency $\omega_{\rm IR}$. It goes to zero when $\mu\to 0$.   We
show that above $\omega_{\rm IR}$, phonons cease to exist. They
are transformed to diffusons. In Section~\ref{diff} we consider
properties of diffusons. The number of diffusing physical
quantities coincides with the number of integrals of motion. In a
closed free mechanical system there are two integrals of motion,
momentum and energy. In Section~\ref{diffmomentum} we investigate
diffusion of momentum. We show that for all masses equal, it is
equivalent to the diffusion of particle displacements since center
of inertia is conserved. The displacement structure factor $S({\bf
q}, \omega)$ coincides well with the structure factor $S_{\rm
rw}({\bf q}, \omega)$ of the random walk on a lattice. We
introduce new additional diffusion coefficient $D_u$ which is {\em
diffusivity of particle displacements}. It is different from the
energy diffusivity  $D(\omega)$ investigated in~\cite{Nature1,
Nature2, Nature3, Nature4, Nature5}. We calculate the correlation
function of particle displacements $C({\bf r}, \omega)$ and radius
of diffuson.  In Section~\ref{ub55p}, we investigate the diffusion
of energy $D(\omega)$ using two different approaches. The first
approach is a direct numerical solution of Newton's equations. In
the second approach the diffusivity is calculated by means of
Edwards and Thouless formula which relates the energy diffusivity
with an infinitesimal change of boundary conditions. Both
approaches give similar results. We show that diffusivity
$D(\omega)$ is independent of frequency in the diffuson range. In
Section~\ref{scaling} we discuss  scaling properties of the model
(their dependence on  parameters $V$ and $\mu$). We show that they
are similar with systems near jamming transition point. In
Section~\ref{disc} we discuss the obtained results and compare
them with experiment.

\section{A random matrix approach}
\label{rma}

In harmonic approximation  vibrational properties of a mechanical
system of $N$ particles are  determined by the dynamical matrix
$M_{ij}=\Phi_{ij}/\sqrt{m_im_j}$, where $\Phi_{ij}$ is the force
constant matrix and $m_i$ are the particle masses. The matrices
$M$ and $\Phi$ are real, symmetric and {\em positively definite}
matrices $N\times N$ ( for simplicity we will consider a scalar
model). The condition of positive definiteness is important. It
ensures mechanical stability of the system.

One can always present every real, symmetric and positive definite
matrix $M$ in the following form~\cite{PositiveDefine,Chalker}
\begin{equation}
    M = AA^T, \quad \mbox{or}\quad M_{ij}=\sum_k A_{ik} A_{jk}.
    \label{we45}
\end{equation}
Here $A$ is some real matrix of a general form (not necessarily
symmetric). And, vice versa, for every real matrix $A$ the product
$AA^T$ is always a positively definite symmetric matrix. Matrices $AA^T$ belong to Wishart ensemble~\cite{Wishart}. The eigenvalue distribution for such kind of large random matrices was firstly investigated in~\cite{Marchenko}.

For a free mechanical system it is necessary to satisfy also
conditions~\cite{maradudin}
\begin{equation}
    \sum_iM_{ij}=\sum_jM_{ij}=0
    \label{s6cv1}
\end{equation}
(for simplicity we consider below all masses $m_i=1$). It ensures
that the potential energy of the system
\begin{equation}
U=\frac{1}{2}\sum\limits_{ij}M_{ij}u_iu_j=-\frac{1}{2}\sum\limits_{i,j<i}M_{ij}(u_i-u_j)^2
\end{equation}
and forces between the particles depend only on the differences of
particle displacements $u_i-u_j$. As a result, the potential
energy and forces are not changed under any translation of the
system as a whole. These conditions are necessary (but not
sufficient) for existence of low frequency acoustic phonon-like
modes in the system. If  conditions (\ref{s6cv1}) are violated, we
have spatially pinned system where propagation of Goldstone modes
(phonons) is not possible.

In structural glasses in many cases (as, for example, in vitreous
silica or amorphous silicon) a mass disorder is not important
and we usually deal with the force constant disorder. It is
related to fluctuations of valence bond lengths and valence bond
angles because of an absence of crystalline ordering. Since
valence forces depend exponentially on the distances between the
atoms, they can experience strong fluctuations. Due to positional
disorder there are also fluctuations of long distance Coulomb
forces in non covalent materials. Thus the force-constant disorder
plays an essential role in glassy dynamics. Therefore, one may
expect that some important properties of glasses can be reproduced
if we take matrix $A$ as a random one.

As soon as the random elastic spring constants $-M_{ij}$ connecting the
particles are fixed, the exact equilibrium particle positions are
no longer important for particle dynamic on a long length scales
much bigger than the interatomic distances. They do not enter to
the dynamical matrix $M$. Therefore, it is reasonable to consider
harmonic {\em lattice models} involving only force constant
disorder.

If disorder is sufficiently strong, then it automatically includes
the coordination number disorder since any weak bonding is
equivalent to a negligibly small interaction between the two
neighbors. Also, if the average coordination number is
sufficiently big, its exact value and, therefore, the type of the
lattice is of no importance as well. The similar considerations
concern polarization of the modes. A vectorial character of
vibrations in real glasses makes the issue to be more complicated.
Many universal properties of glasses (as, for example, the thermal
conductivity) are not related to polarization of the modes.
Therefore, for better understanding of the physics involved it is
often instructive to exploit scalar models. In these models the problem of
zero frequency modes existing in vectorial isostatic lattices~\cite{mao}
does not exist~\cite{schirm}. For these reasons,
different scalar models were successfully used in glassy physics
in the past~\cite{schirm, parisi1, grigera, bunde}.

Due to the reasons mentioned above we, as in~\cite{our2}, consider
the case of a simple cubic lattice with $N$ particles and lattice
constant $a_0=1$. Each particle has its unique integer index $i$
which takes values from $1$ to $N$. We construct the random matrix
$A$ as follows. The non-diagonal elements $A_{ij}$ (for $i\ne j$)
we take as independent random numbers from Gaussian distribution
with zero mean $\left<A_{ij}\right>=0$ and unit variance
$\left<A^2_{ij}\right>=V^2=1$ if $i$-th and $j$-th particles are
nearest neighbors. For each particle in a simple cubic lattice
there are six nearest neighbors. As a result, for given $i$ we
have 6 non zero non-diagonal elements $A_{ij}$ for matrix $A$.
Non-diagonal elements $A_{ij}$ and $A_{ji}$ are statistically
independent from each other (matrix $A$ is non-symmetric). All
other non-diagonal elements (for non-nearest neighbors)
$A_{ij}=0$. To ensure the property (\ref{s6cv1}) the diagonal
elements $A_{ii}$ are calculated as follows
\begin{equation}
    A_{ii}= -\sum\limits_{j\ne i} A_{ji} .
    \label{67vb}
\end{equation}
Then, according to Eq.~(\ref{we45}), the Eq.~(\ref{s6cv1}) will be
also met.

\begin{figure}[h!]
     \includegraphics[scale=0.3]{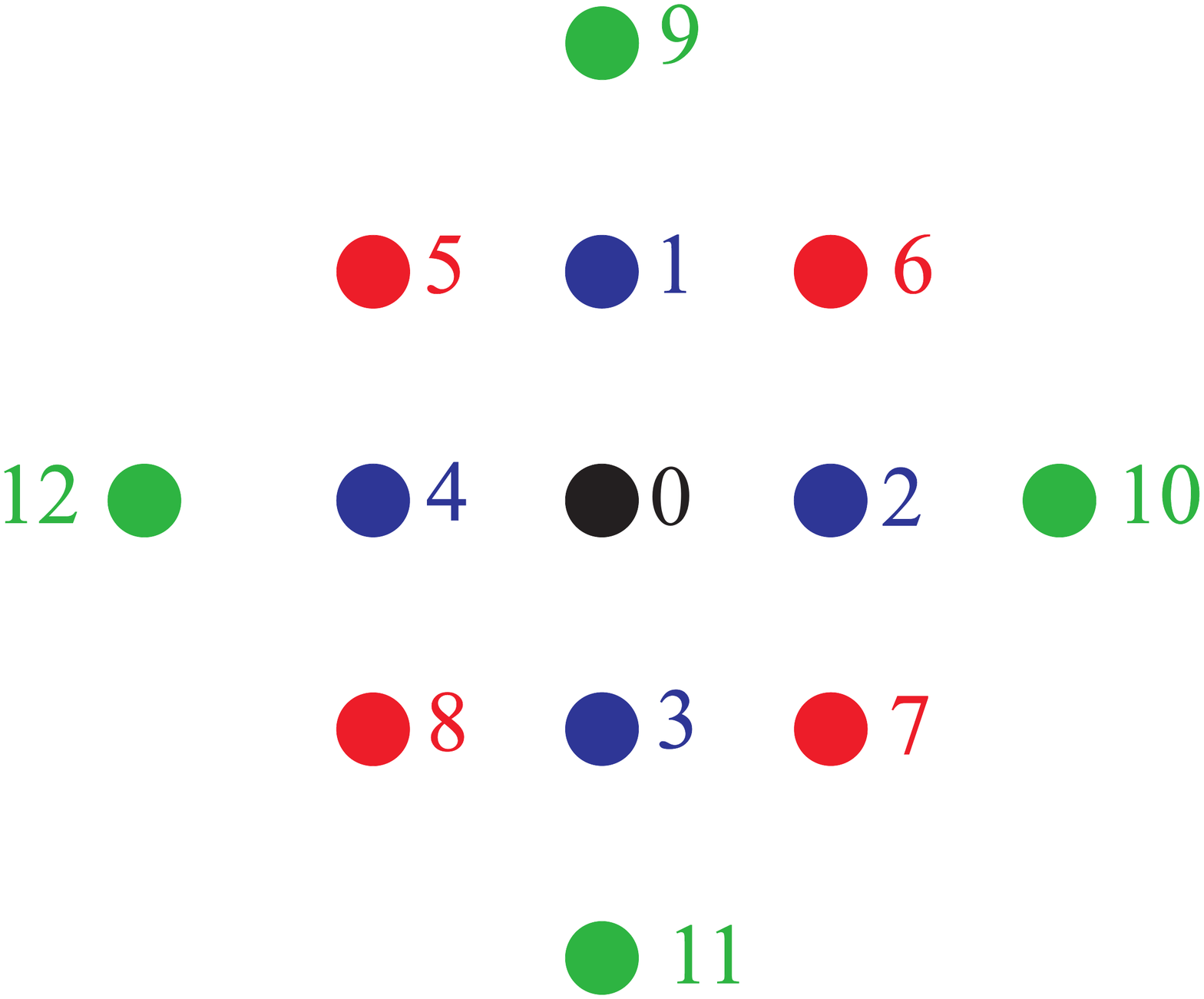}
    \caption{Structure of the dynamical matrix $M$ in $2d$ case. Particles 1-12 interact with the central black particle.}
    \label{fig:dyn_matrix}
\end{figure}
The matrix $M$ is then constructed according to Eq.~(\ref{we45}).
As was shown in~\cite{our1}, in a $3d$ simple cubic lattice each
particle is connected by elastic springs with $24$ neighbors. The
elastic spring constants are random and can be either positive or
negative. A negative spring by definition is a spring which
expands after initial stretching and shrinks after initial
contraction. The effect of negative spring constants on atomic
vibrations was discussed in different papers~\cite{schirm, resmod,
surface, frenkel, electron, parisi1, taraskin3}.

To elucidate where this coordination number 24 in $3d$ case comes
from, let us first consider  as example a $2d$ simple square
lattice. In this case each particle interacts with 12 neighbors
shown on Fig.~\ref{fig:dyn_matrix}.  In accordance to
Eq.~(\ref{we45}), matrix elements connecting central black
particle with its 4 nearest neighbors are of the type
\begin{equation}
M_{01}=\sum\limits_k A_{0k}A_{1k}=A_{00}A_{10}+A_{01}A_{11}.
\label{z8bn}
\end{equation}
From Eq.~(\ref{67vb}) it follows that diagonal elements meet the
following relations
\begin{equation}
A_{00}=-(A_{10}+A_{20}+A_{30}+A_{40}), \label{sc45}
\end{equation}
\begin{equation}
A_{11}=-(A_{91}+A_{51}+A_{61}+A_{01}). \label{sc34}
\end{equation}
One has to insert them in Eq.~(\ref{z8bn})
\begin{eqnarray}
M_{01}= &-& A_{10}^2 - A_{01}^2
        - A_{10}(A_{20}+A_{30}+A_{40})-\nonumber \\
        &-& A_{01}(A_{91}+A_{51}+A_{61}) .
\label{h7cv}
\end{eqnarray}
Since averaged values $\left<A_{i\ne j}\right>=0$ and different
non-diagonal matrix elements $A_{ij}$ are statistically
independent from each other, the average value $\left<M_{01}
\right>$ is determined by the first two quadratic terms in
Eq.~(\ref{h7cv}). As a result, it is non-zero and negative. It
corresponds to positive average elastic spring $k_{01}=-M_{01}$
between particles 0 and 1
\begin{equation}
\left<k_{01}\right>=-\left<M_{01}\right>=\left<A_{10}^2\right>+\left<A_{01}^2\right>=2.
\label{c6x4}
\end{equation}
Though, according to Gaussian distribution of $A_{i\ne k}$, the
spring constant $k_{01}$ can take negative values as well. All the
aforesaid  is valid for other nearest neighbor matrix elements
$M_{02}$, $M_{03}$ and $M_{04}$.

The next nearest neighbor matrix elements are given by
\begin{equation}
M_{05}=\sum\limits_k A_{0k}A_{5k}=A_{01}A_{51}+A_{04}A_{54} ,
\label{5xwe}
\end{equation}
\begin{equation}
M_{09}=\sum\limits_k A_{0k}A_{9k}=A_{01}A_{91} . \label{3vcb}
\end{equation}
It is easy to see that the average values of $\left<M_{05}\right>$
and $\left<M_{09}\right>$ and corresponding average elastic
springs $\left<k_{05}\right>$ and $\left<k_{09}\right>$ are zero.
So the next nearest neighbor springs can be  either positive or
negative with equal probability. The same is valid for 6 other
next nearest neighbor matrix elements $M_{06}$, $M_{07}$, $M_{08}$
and $M_{0,10}$ $M_{0,11}$, $M_{0,12}$.

\begin{figure}[!h]
    \includegraphics[scale=0.4]{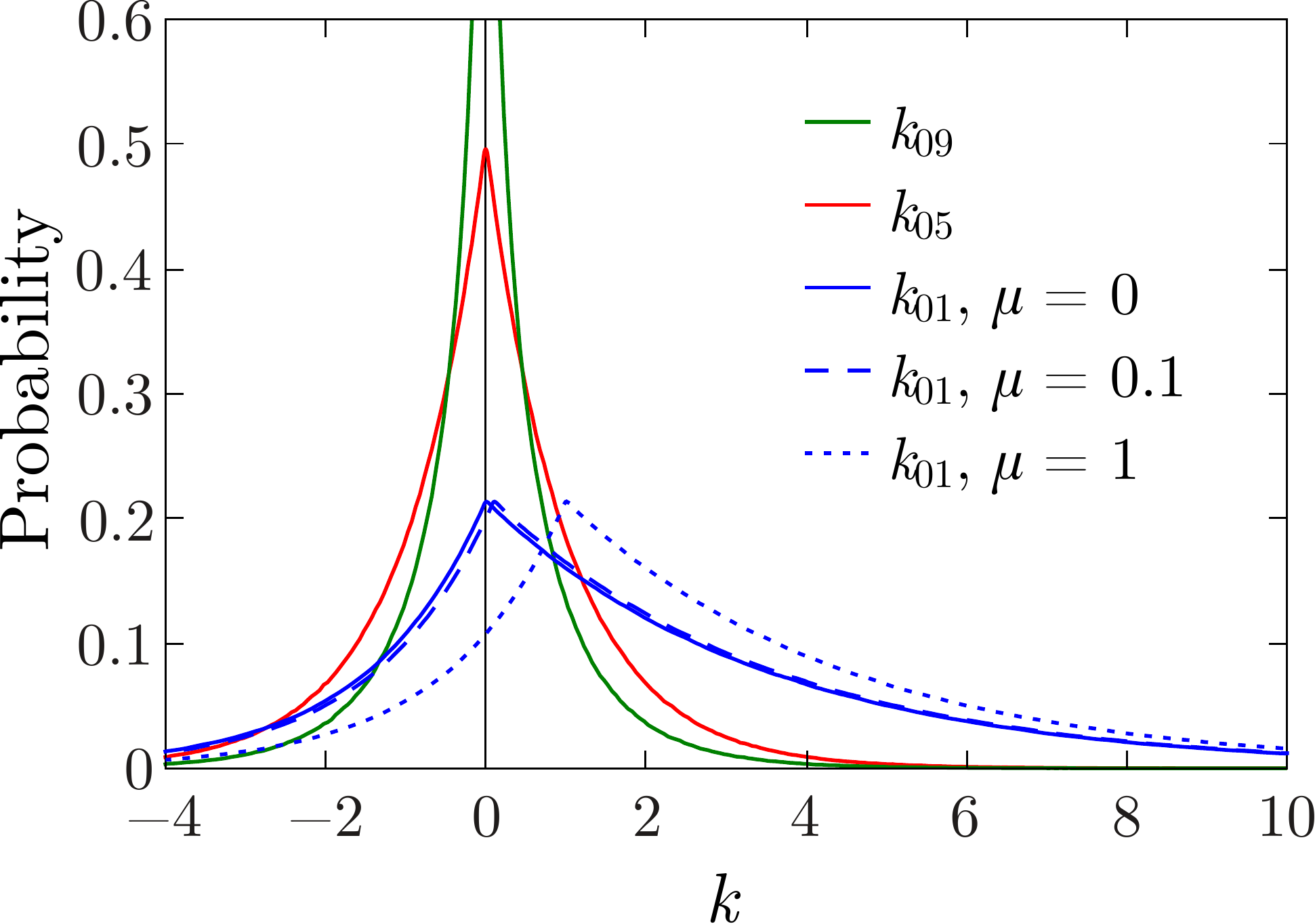}
    \caption{Distributions of random elastic spring constants in $3d$ simple cubic lattice.}
    \label{fig:kDistr}
\end{figure}
In 3d case for simple cubic lattice there are 6 springs of the
type $M_{01}$, 12 springs of the type $M_{05}$ and 6 springs of
the type $M_{09}$. As a result all together we have 24
particles interacting with the central black particle. All these
24 spring constants can be either positive or negative but to
ensure the mechanical stability of the whole system they are
correlated with each other in a rather tangly way.

Distributions of different spring constants are shown on
Fig.~\ref{fig:kDistr}. The distribution of $k_{01}$ is asymmetric
with positive mean value. The distributions of $k_{05}$ and
$k_{09}$ are even (with zero average value) and for $k_{09}$ are given by zeroth-order
Macdonald function~\cite{our1} which logarithmically diverges at
$k=0$. The resulting distribution of all spring constants was
calculated numerically in~\cite{our2}. The number of negative
springs was found to be about 45\%. One can find a similarity
between our spring constant distributions and dynamical matrix
element distributions obtained in~\cite{taras} for IC-glass,
in~\cite{huang} for  simple fluid with short-ranged interactions
(see Fig.~1 in these papers), and in~\cite{christie0}  for
realistic model of amorphous silicon (see Figs~2.12, 2.13). Though
it is difficult to compare our scalar model with vector models
analyzed  in~\cite{taras, huang, christie0}.

Concluding this part, we can easily include into consideration the
next neighbor shell for matrix $A$. Then, in addition to the
previous case, the matrix elements of the type $A_{05}$  should be
taken into account. As a result the coordination number for matrix $M$ in simple cubic lattice increases up to 90.
Just opposite, applying some additional
constraints, we can reduce the coordination number from 24 to smaller numbers or make it
fluctuating quantity, etc. We have checked that all these
modifications can lead to quantitative changes but do not change
qualitatively the main results of the paper. Therefore, we will
restrict our consideration by the simplest case outlined before.

\begin{figure}[!h]
    \includegraphics[totalheight=5cm,keepaspectratio]{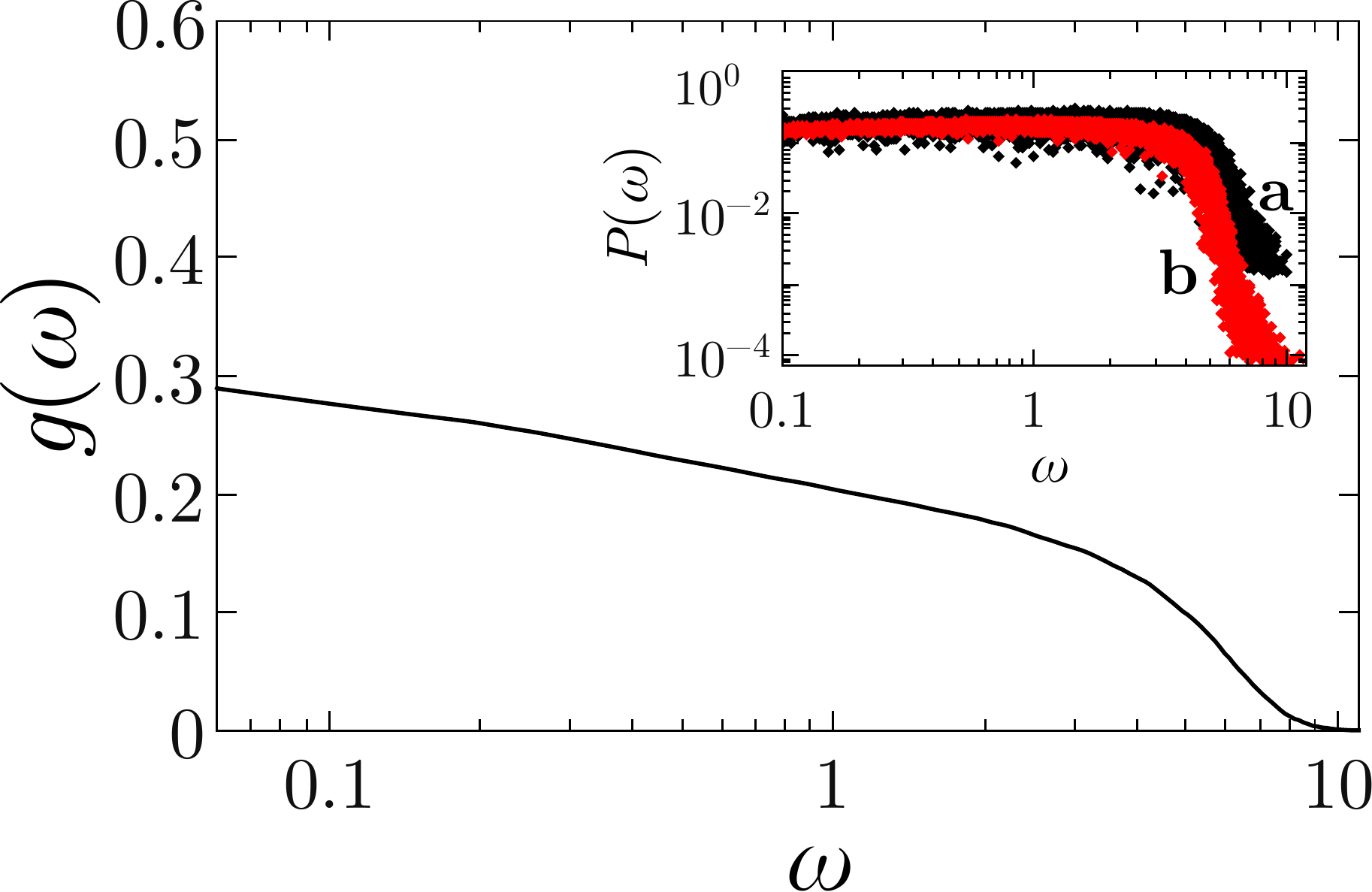}
    \caption{The normalized DOS $g(\omega)$ for random matrix $M=AA^T$ built on a simple cubic lattice
    with $N=20\times 20\times 20$ particles and averaged over $1000$ realizations. Inset: The participation
    ratio $P(\omega)$ for $N=10^3$ (a) and $N=27^3$ (b) for one realization.}
    \label{fig:Dos1}
\end{figure}
Fig.~\ref{fig:Dos1} shows the normalized density of vibrational
states (DOS) $g(\omega)$ of matrix $M=AA^T$ in $3d$ simple cubic
lattice. The periodic boundary conditions were used. As follows
from the figure, the spectrum is gapless i.e. $g(\omega)$ is
nonzero at $\omega\to 0$. In spite of the fact that the conditions
(\ref{s6cv1}) are fulfilled,  we do not see the expected phonon
modes with their DOS $g_{\rm ph}(\omega)\propto\omega^2$ for
$\omega\to 0$. It means that phonons as plane wave excitations
cannot propagate through the lattice. This result is not changed
qualitatively ~\cite{our2} neither by including next neighbor
shells to build matrix $A$ nor by switch to vector model.

As was
shown in~\cite{our2}, such a behavior of DOS at $\omega\to 0$ is
related to the fact that the affine assumptions are violated and
the macroscopic elasticity theory becomes inapplicable in this
case. The average value of the static Young modulus of the lattice
$E\propto 1/N$. Therefore, in the thermodynamic limit
($N\to\infty$) $E\to 0$. As a result, the rigidity of the lattice
and sound velocity are also tend to zero. This unusual behavior is
due to a presence of high concentration of negative springs (45\%)
in the lattice which makes it to be extremely soft.

\begin{figure}[h!]
    \centerline{\includegraphics[height=65mm, bb = 0 0 400 400]{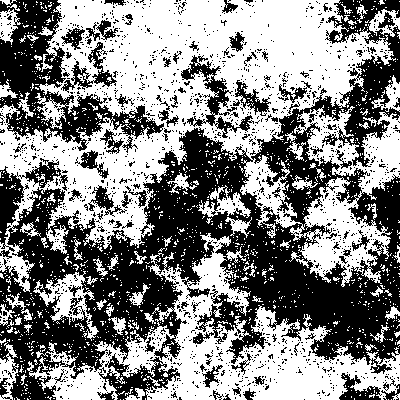}}
    \caption{The spacial eigenmode structure of random matrix $M=AA^T$ for the lowest
    frequency $\omega_{\rm min}$ in two dimensional square lattice $400\times 400$.}
    \label{fig:fractal}
\end{figure}

To determine whether vibrational  modes are localized or
delocalized, we have calculated the participation ratio
\begin{equation}
P(\omega)=\left[N\sum\limits_{i=1}^Ne_i^4(\omega) \right]^{-1}.
\label{r6gh}
\end{equation}
Here $e_i(\omega)$ is $i$-th particle projection of the normalized
eigenvector with frequency $\omega$. As one can see from the inset
of Fig.~\ref{fig:Dos1}, all modes with exception of small high
frequency part are {\em delocalized}. They have $P(\omega)\approx
0.2$ which is independent of the system size.  This value is close
to the theoretical value $1/3$ for Porter-Thomas distribution of
$e_i^2(\omega)$~\cite{porter,our1}. We have verified also that the
level spacing distribution obeys the Wigner-Dyson
statistics~\cite{our1}. It also indicates the mode delocalization.
As we will show in Section~\ref{diff}, all these delocalized
gapless vibrational modes can be identified as diffusons. They
spread in the lattice by means of diffusion.

To elucidate a spacial structure of the eigenmodes for matrix
$M=AA^T$ we considered as an example a two dimensional square
lattice with $N=400\times 400$ particles and calculated
eigenvector $e_i(\omega_{\rm min})$ ($i=1, 2, ... , N$)  for the
lowest frequency $\omega_{\rm min}$ in the system. The result is
shown on Fig.~\ref{fig:fractal}. Particles with positive and
negative displacements are shown by white and black dots
correspondingly. As one can see from the figure, the mode is
delocalized. Its spatial structure is random (fractal) and has
nothing to do with a plane wave. Similar  picture takes place in a
3d case.

\section{Phonons}
\label{phonons}

To introduce phonons into the picture we should have  finite
rigidity of the lattice. The rigidity can be introduced by
different means. Since a sum of positive definite matrices is a
positive definite matrix, then simplest possibility is to add to the
random matrix $AA^T$ a ``crystalline part''~\cite{our2}
\begin{equation}
    M=AA^T+\mu M_0.
\label{rt56e}
\end{equation}
Here $A$ is the same random matrix built on a $3d$ simple
cubic lattice with $a_0=1$ as in the previous Section. Matrix
$M_0$ is a positively definite crystal dynamical matrix for the
same lattice with unit masses, and all spring constants (between
the nearest neighbors) equal to unity. As was shown in~\cite{our2}
the tune parameter $\mu\geqslant 0$ controls the rigidity of the
lattice.

Adding the regular part $\mu M_0$, changes the distribution of
spring constants $k_{01}$ between the nearest neighbors, as shown on
Fig.~\ref{fig:kDistr}. The average value is equal to
$\left<k_{01}\right>=2+\mu$. At small values of $\mu\ll 1$ the
change  is negligible. The distribution mainly consists from
strongly fluctuating part $AA^T$ (compare the distributions of
$k_{01}$ for $\mu=0$ and $\mu=0.1$). Therefore, it is not obvious
at all that such small perturbation is able to introduce a finite
rigidity and phonons into the system. A strong
scattering of phonons by diffusons may leave the diffuson spectrum
unchanged.

Also we can consider elastic springs in matrix $\mu M_0$ to
be fluctuating quantities, somehow  distributed in the closed
interval $[0, \mu]$. Otherwise we can cut out a big amount of
springs $\mu$ from the lattice, so the phonons cease to exist in
the term $\mu M_0$ at all (see Section~\ref{cutlat}). Another
nontrivial possibility is shown in Section~\ref{superposition}.
In the paper we limit ourselves by the most simple case described
by Eq.~(\ref{rt56e}).

\begin{figure}[!h]
    \includegraphics[scale=0.4]{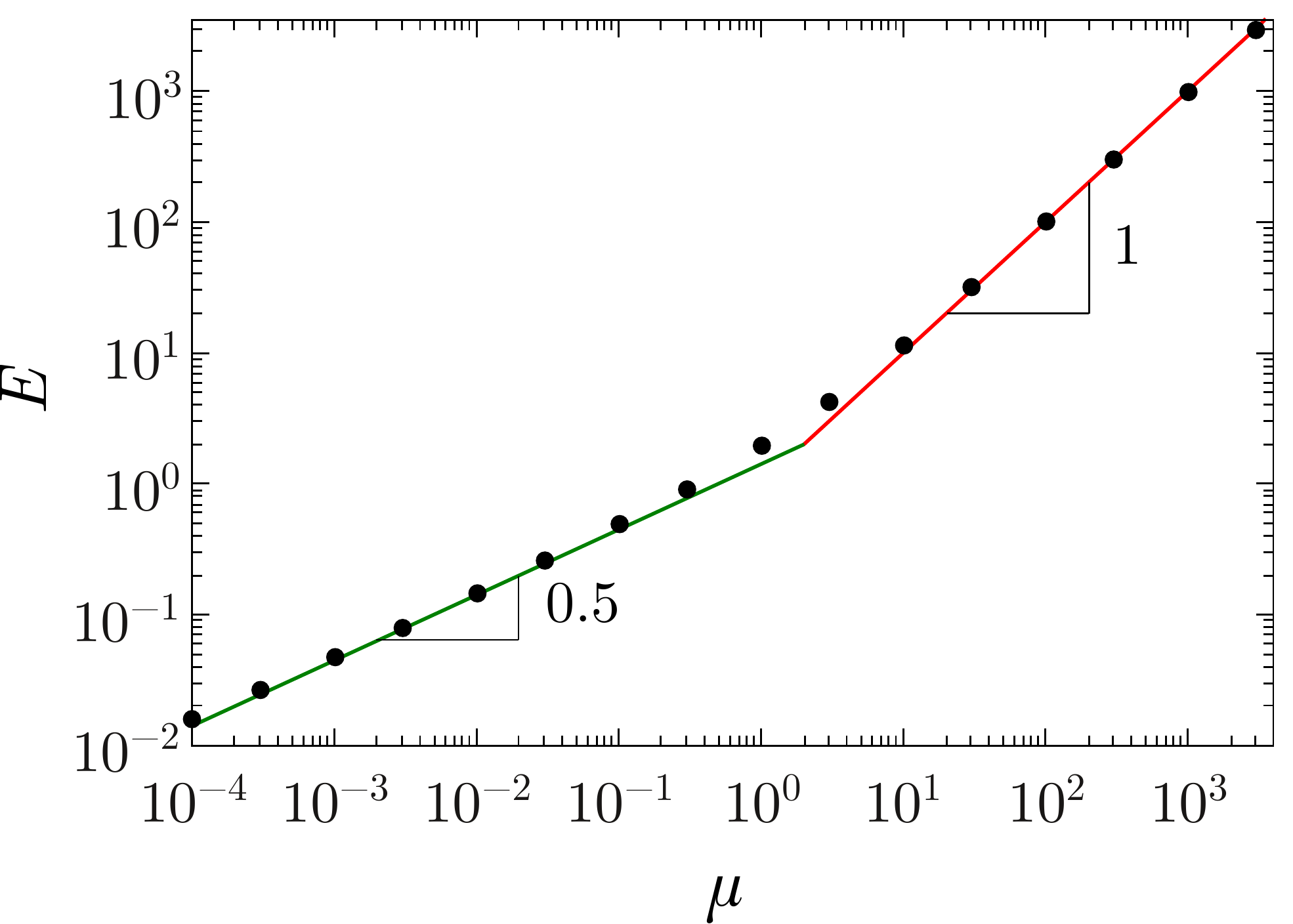}
    \caption{Young modulus $E$ as a function of $\mu$ for dynamical matrix $M=AA^T+\mu M_0$
    built on a cubic lattice with $N=100\times 100\times 100$ particles (one realization).
    Black dots are calculated values, the line is the best least-square fit.}
    \label{fig:Emu}
\end{figure}
To find the rigidity (as a function of $\mu$), we calculated
numerically the Young modulus $E$ of the lattice with dynamical
matrix given by Eq.~(\ref{rt56e}) for $\mu\ne 0$. In modeling of
amorphous solids, the standard method to do that is to use
Irving-Kirkwood stress tensor formula~\cite{irving}. However, it
is difficult to implement this procedure in our case of strong
local fluctuations of elastic springs when microscopic
displacement field $u({\bf r})$ is not a differentiable function
of atomic positions~\cite{our2}.

Therefore, to avoid these difficulties we, as in~\cite{our2}, used
a direct numerical method. We took a very big cubic sample with
$N=L\times L\times L=10^6$ particles and side $L-1$ to reduce
fluctuations and possible non-affine response. According to the
standard textbook formula of the macroscopic elasticity theory
(see Eq.~5.2 in~\cite{Landau}), Young modulus is given by
$E=\sigma_{zz}/u_{zz}$. Here $\sigma_{zz}$ is the stress, and
$u_{zz}$ is the strain. The component $u_{zz}$ gives the relative
lengthening of the sample $\Delta L/(L-1)$. Then we fixed the
strain and calculated the stress $\sigma_{zz}$.

For that we fixed particles on the left hand side of our cubic
sample  and displaced all particles on the opposite (right hand)
side  by the unit distance $\Delta L=1$. Since Newton equations
are linear, the final result is independent of the value of the
step strain used. In other two directions we used the periodic
boundary conditions. Then, solving the system of linear Newton
equations, we found the new equilibrium positions of all other
particles in the sample and calculated restoring forces $f_i$
acting on the displaced particles on the right boundary. Due to
randomness of the elastic bonds, the restoring forces are also
random. Let $\bar f$ be the average restoring force. Then,  by
definition,  the stress $\sigma_{zz}=\sum_if_i/L^2=\bar{f}$ and the
Young modulus $E$ can be calculated as follows
\begin{equation}
E=\frac{\bar{f}(L-1)}{\Delta L} . \label{d6vg}
\end{equation}
To avoid confusion, we remind that we are using here a scalar
version of the elasticity theory. Therefore, all forces in the
lattice are parallel (or antiparallel) to the particle
displacements.

The results of these calculations are shown on Fig.~\ref{fig:Emu}
for cubic sample with $N=10^6$ particles~\cite{fluct}. As we can
see from the fit, the Young modulus has a following dependence on
$\mu$:
\begin{eqnarray}
E &=& \mu, \quad \mu\gg 1, \label{a6df}\\
E &=& 1.5\sqrt{\mu}, \quad \mu \ll 1 . \label{x6sd}
\end{eqnarray}
As a result, for $\mu\gg 1$ we have a usual crystal, where
disorder is relatively small and relation (\ref{a6df}) is obvious.
For $\mu\ll 1$ the force constant disorder is strong. The
fluctuations of the nondiagonal matrix elements $M_{i\ne j}$ are
much bigger than the averaged values~\cite{our1, our2}. In this
case Young modulus $E\propto\sqrt{\mu}$. It is much bigger than
the crystal result (\ref{a6df}). Strong fluctuations of the
positive and negative elastic springs which in average almost
compensate each other make the lattice much more rigid than in the
case of crystal. Therefore for $\mu\ll 1$ one can not consider our
lattice as a simple superposition of two systems $AA^T$ and $\mu
M_0$. The origin of this behavior $E\propto\sqrt{\mu}$ is unclear and it should be
elucidated in further work (see also Section~\ref{scaling}). But
below we will support our numerical findings by calculation of the
sound velocity and of the phonon density of states (for small
$\omega$) and by a comparison of the latter with total DOS
calculated numerically for matrix (\ref{rt56e}). Below in this
paper we will consider the case of strong and moderate force
constant disorder when $0\leqslant\mu\leqslant 1$.

\begin{figure}[!h]
    \includegraphics[scale=0.4,keepaspectratio]{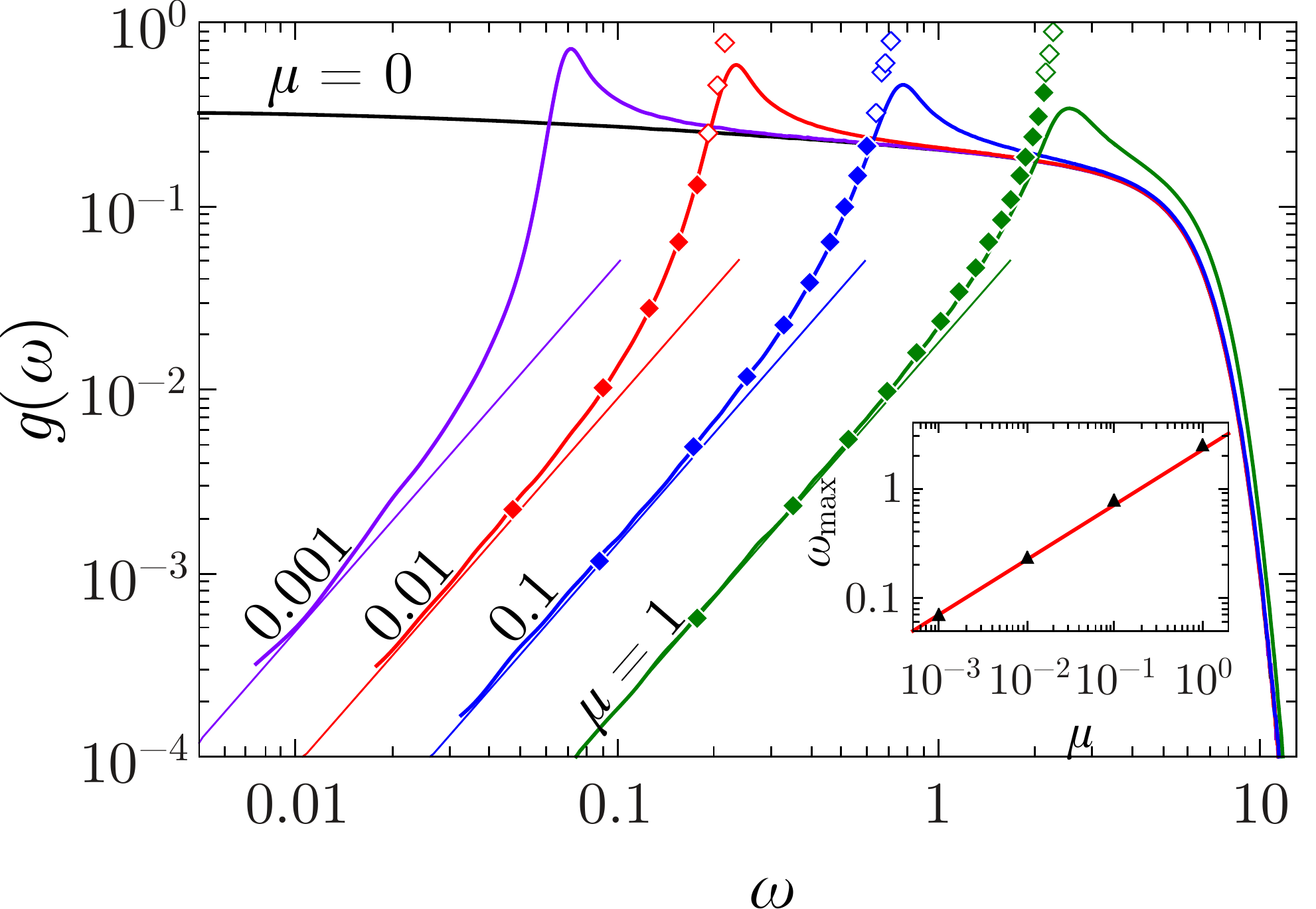}
    \caption{The normalized DOS $g(\omega)$ for dynamical matrix $M=AA^T+\mu M_0$ and four different
    $\mu$ (0, 0.001, 0.01, 0.1, 1) calculated with precise numerical KPM  solution for cubic
    lattice with $N=200^3$ (full lines). Straight lines correspond to Eq.~(\ref{st56}) with sound
    velocity $v=\sqrt{E}$. Filled and open diamonds correspond to phonon contribution to the DOS below and
    above the Ioffe-Regel crossover frequency $\omega_{\rm IR}$ correspondingly (see further text for details).
    Inset: dependence $\omega_{\rm max}(\mu)\propto\sqrt{\mu}$.}
    \label{fig:gwmu}
\end{figure}

To calculate the phonon contribution to the DOS at small $\omega$,
we need to know the sound velocity $v$ at zero frequency. It is
related to the Young modulus in a  standard way:
\begin{equation}
v=\sqrt{E} \label{s7cvbf}
\end{equation}
(since all particle masses $m_i=1$ and lattice constant $a_0=1$).
Then for the phonon DOS (in the scalar model) we have
\begin{equation}
g_{\rm ph}(\omega)=\frac{1}{2\pi^2}\frac{\omega^2}{v^3}.
\label{st56}
\end{equation}

The total DOS $g(\omega)$, normalized to unity and calculated
numerically by the kernel polynomial method (KPM)~\cite{kpm,
kpmreview} for dynamical matrix (\ref{rt56e}) and different values
of $\mu$, is shown on Fig.~\ref{fig:gwmu}. We see from the figure
that for $\mu\ne 0$ the DOS at low enough frequencies is
proportional to $\omega^2$ which corresponds to acoustical phonon
excitations. Thus, introducing finite values of $\mu$, we open up a
soft {\em phonon gap} in the gapless diffuson spectrum, existing at $\mu=0$.
The DOS in the gap, as we will show in the paper, is built
by acoustic phonon-like modes and at low frequencies goes to zero
as $g(\omega)\propto\omega^2$. The term {\em phonon gap} is motivated
since, if conditions (\ref{s6cv1}) are violated, then addition $\mu
M_0$ to random matrix $AA^T$ opens a {\em hard gap} in the gapless
vibrational spectrum (see Fig.~\ref{fig:hardgap} below).
Just above this gap the DOS has a sharp maximum at
frequency $\omega_{\rm max}$ which we will identify with the width
of the gap. As follows from the figure, the maximum frequency for
$\mu\ll 1$ increases as $\omega_{\rm max}\propto\sqrt{\mu}$. Above
the maximum the vibrational excitations remain to be diffusons
(see Section~\ref{diff}).

One can try to explain  the dependence $\omega_{\rm max}\propto\sqrt{\mu}$
as follows. In the absence of random part $AA^T$ the
dynamical matrix $M$ is determined by the crystalline part $\mu
M_0$ only.  Then (for simple cubic lattice) we have well defined
phonon modes with dispersion law
\begin{equation}
\omega^2_{\rm cryst}=4\mu\left(\sin^2\frac{q_x}{2} +
\sin^2\frac{q_y}{2}+ \sin^2\frac{q_z}{2}\right).
\end{equation}
The maximum frequency in this case is equal to $\omega_{\rm max,\,
cryst}=2\sqrt{3\mu}\propto\sqrt{\mu}$ which qualitatively (but not
quantitatively) explains aforesaid dependence $\propto\sqrt{\mu}$.
However the sound velocity in this  pure crystallyne lattice case
$v_{\rm cryst}=\sqrt{\mu}$.  Though according to
Eqs.~(\ref{s7cvbf}, \ref{x6sd})  $v\propto\mu^{1/4}$ for
$M=AA^T+\mu M_0$ what is much bigger then $\sqrt{\mu}$  for small
values of $\mu\ll 1$. It means that simple superposition approach
does not work in this case and physical picture is more
complicated. As we will show in Section~\ref{scaling} the Young
modulus $E$ depends also on the amplitude of the random part $AA^T$.

Since the DOS $g(\omega)$ is normalized to unity for all values of
$\mu$, we conclude from Fig.~\ref{fig:gwmu}  (comparing the DOS
for $\mu\ne 0$ with DOS for $\mu=0$) that vibrations corresponding
to the maximum for $\mu\ne 0$ were pushed out from the region of
small frequencies $\omega<\omega_{\rm max}$ for $\mu=0$.  We see
also from the figure that, after initial $\omega^2$ dependence,
the DOS for $\mu\ne 0$ increases much faster than $\omega^2$. It
is a clear signature of the presence of the boson peak in our
disordered lattice. As we will show further (see Table I), the
frequency $\omega_{\rm max}$ is correlated with position of the
boson peak $\omega_b$ (the maximum in the reduced DOS
$g(\omega)/\omega^2$). Therefore appearance of the boson peak in
disordered systems is not necessarily related to the acoustic van
Hove singularity in crystals as was proposed
recently~\cite{schirm, taraskin3, vanHove2}.

The straight lines on the Fig.~\ref{fig:gwmu} correspond to the
phonon DOS $g_{\rm ph}(\omega)$ determined by Eq.~(\ref{st56})
with the sound velocity $v=\sqrt{E}$ and $E$ calculated from
Fig.~\ref{fig:Emu}. One can see a good agreement of the total
$g(\omega)$ at low frequencies with the phonon contribution
$g_{\rm ph}(\omega)$. From that we can conclude that at least the
low frequency excitations in the phonon gap are the usual
long-wave acoustical phonons. However, actually, as we will show
further, nearly all excitations in the gap up to the frequencies
close to $\omega_{\rm max}$ correspond to phonons, but with a
nonlinear dispersion law.

\begin{figure}[h!]
     \includegraphics[scale=0.4]{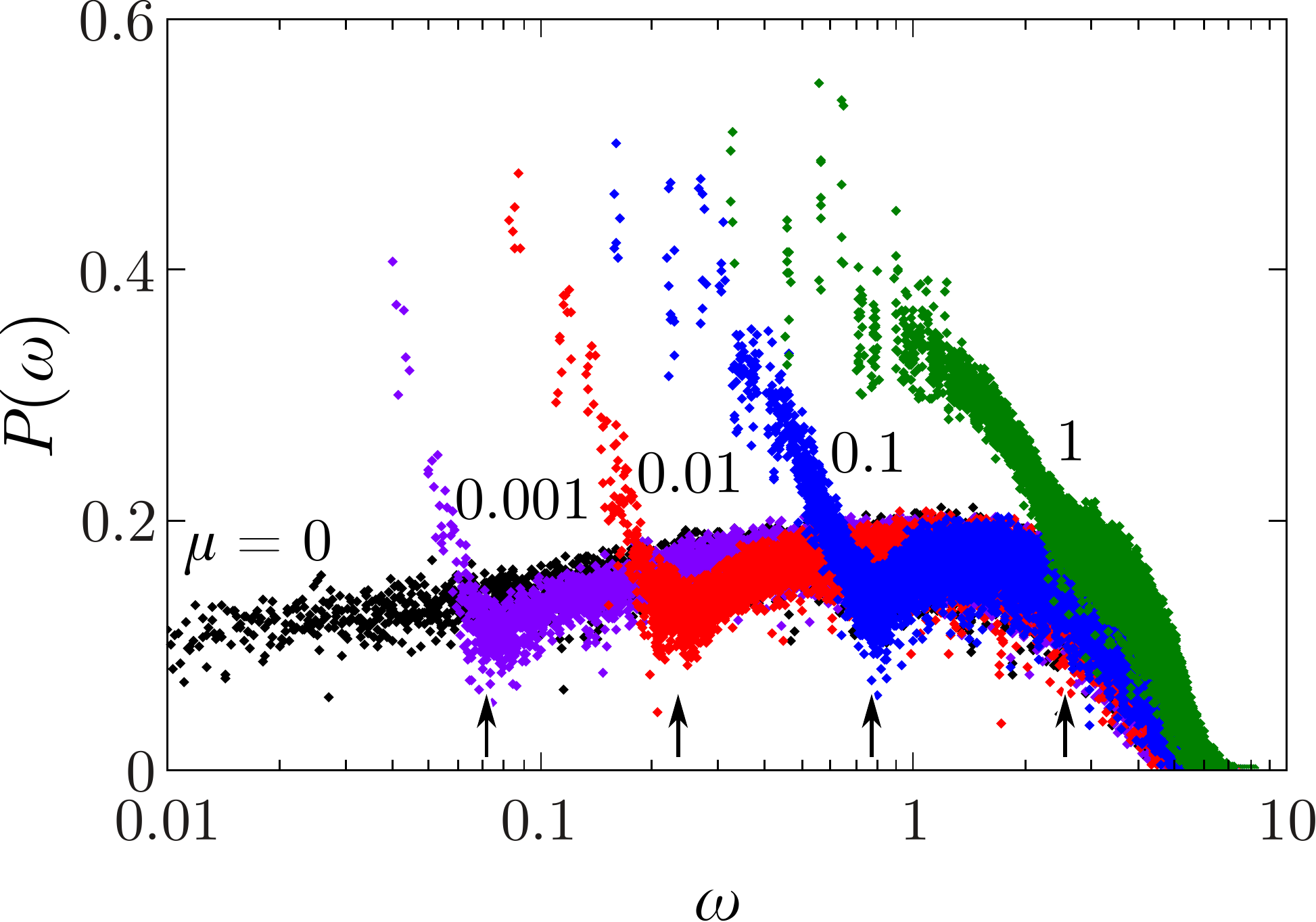}
     \caption{Participation ratio for different $\mu$ as a function of $\omega$ for $N=27^3$
     (one realization). The arrows indicate positions of $\omega_{\rm max}$ in $g(\omega)$
     for corresponding values of $\mu$ (see Fig.~\ref{fig:gwmu}).}
     \label{fig:PR}
\end{figure}

This conclusion is supported by calculations of the participation
ratio $P(\omega)$. It is shown in Fig.~\ref{fig:PR} for various
values of $\mu$. For $\mu\ne 0$, one can clearly distinguish  in
the function $P(\omega)$ a presence of the two different frequency
regions. As follows from Fig.~\ref{fig:gwmu}, the low frequency
part (below $\omega_{\rm max}$) corresponds to the phonons. In
this range the participation ratio increases with decreasing
frequency.  It is related to increase of the phonon mean free path
$l(\omega)$ as $\omega\to 0$ (see Fig.~\ref{fig:IR}). In the high
frequency part (above $\omega_{\rm max}$) $P(\omega)$ is
approximately independent of the frequency and coincides with
participation ratio for $\mu=0$. As we will show in
Section~\ref{diff} this range corresponds to diffusons. A similar rise of
the participation ratio with decreasing frequency was found
recently in 2d Lennard-Jones glasses~\cite{tanguy} (see Fig.~1b of
this paper).

\begin{figure}[!h]
    \includegraphics[scale=0.4,keepaspectratio]{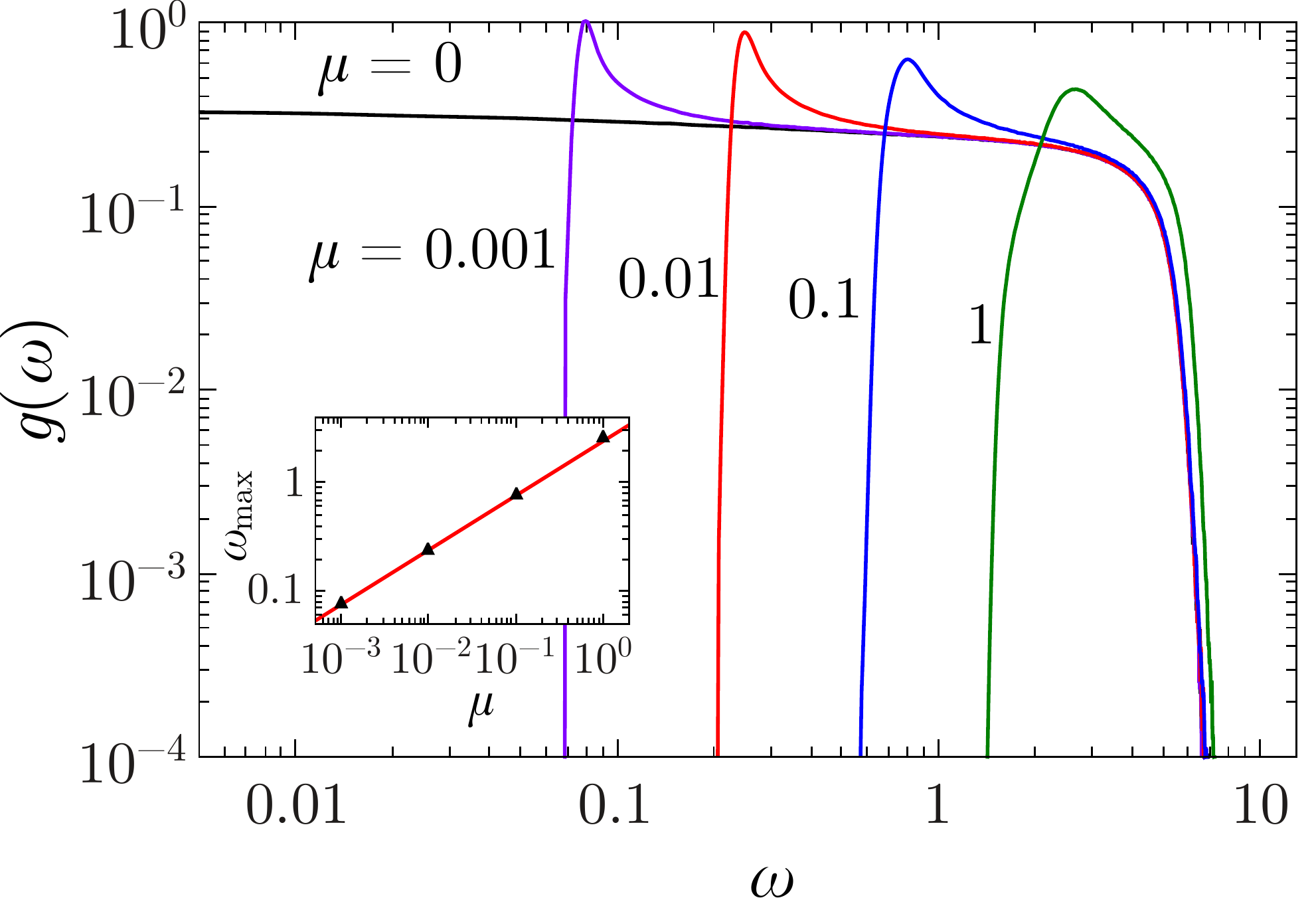}
    \caption{The normalized DOS $g(\omega)$ for dynamical matrix $M=AA^T+\mu M_0$
    and different $\mu$ (0, 0.001, 0.01, 0.1, 1) calculated with precise numerical KPM
    solution for cubic lattice with $N=200^3$ (full lines). The conditions (\ref{s6cv1}) are violated.
    Inset: dependence $\omega_{\rm max}(\mu)\propto\sqrt{\mu}$.}
    \label{fig:hardgap}
\end{figure}
It is important to emphasize that for existence of the acoustical
phonon excitations the conditions (\ref{s6cv1}) are crucial. If
they are not obeyed, then, instead of soft phonon gap in the
vibrational spectrum shown on Fig.~\ref{fig:gwmu}, we have  a hard
gap shown on Fig.~\ref{fig:hardgap}. Inside the hard gap there are
no vibrations at all. The dynamical matrix $M$ in this case was
taken in the same form (\ref{rt56e}). But diagonal elements
$A_{ii}$ of the matrix $A$  were taken  as independent Gaussian
random variables with average $\left<A_{ii}\right>=0$ and unite
variance $\left<A^2_{ii}\right>=1$. As a result the condition
(\ref{67vb}) (and therefore (\ref{s6cv1})) was violated and we
have got a spatially pinned lattice where low frequency acoustical
phonon modes cannot exist. However, the width of the hard gap in
this case has the same $\mu$ dependence as the width of the phonon
gap, $\omega_{\rm max}\propto\sqrt{\mu}$.

To find the phonon dispersion curve (dependence of the phonon
frequency $\omega$ on the wave vector $\bf q$) and phonon mean
free path $l(\omega)$ we should calculate space and time Fourier
transform of the particle displacement field $u({\bf r}, t)$. For
that we ascribed to all the particles at the initial moment $t=0$
random displacements $u({\bf r}, 0)$ (from Gaussian distribution
with zero mean  and unit variance) and zero velocities. Then,
numerically solving Newton equations (with all masses $m_i=1$) we
analyzed the particle dynamics at $t\ne 0$. In calculations we
used Runge-Kutta-4 method with sufficiently small time step
$\Delta t=0.01$.  We have checked that in this case the total
energy of the system is conserved over the whole investigated time
interval $T$ with relative precision $10^{-7}$ without use of any
damping. The calculated values of particle displacements also have
relative precision higher then $10^{-7}$ (we compared the results
with time step $\Delta t=0.01$ with results obtained with two
times smaller time step $\Delta t=0.005$).

Let $u({\bf r}_i, t)$ be the $i$-th particle displacement as a
function of particle coordinate ${\bf r}_i$ and time t. We define
the displacement structure factor (DSF) of the displacement field
as follows
\begin{equation}
S({\bf q}, \omega) =
\frac{2}{NT}\left|\sum\limits_{i=1}^Ne^{-i{\bf q}{\bf
r}_i}\int\limits_0^{T} u({\bf r}_i, t) e^{i\omega t}dt \right|^2 .
\label{45gto}
\end{equation}
For better frequency resolution, the upper time limit $T$ was
taken sufficiently large ($T=3000$), while the integration time
step was chosen as $\Delta t=0.01$. Since vectors ${\bf r}_i$ in a
cubic lattice  are discrete, the wave vectors ${\bf q}\equiv {\bf
q}_n$ are also discrete and are defined on the corresponding
reciprocal lattice. For example, for cubic sample $L\times L\times
L$ and ${\bf q}\parallel \left<100\right>$ direction we have
$q_n=2\pi n/L$ where integer numbers $n$ are $-L/2 \leqslant n
\leqslant L/2$.

One can show (see Section~\ref{DSF}) that definition (\ref{45gto})
is equivalent to the usual expression
\begin{equation}
    S({\bf q}, \omega) = \frac{\pi}{N}\sum\limits_{j=1}^N\Big|\sum\limits_{i=1}^Ne_i(\omega_j)e^{-i{\bf q}{\bf r}_i}\Big|^2\delta(\omega-\omega_j) .
    \label{eq:sqw_def2}
\end{equation}
Here $e_i(\omega_j)$ --- is $i$-th component of the eigenvector of
the dynamical matrix $M$ corresponding to $i$-th particle and
eigenfrequency $\omega_j$~\cite{maradudin}. The normalized density
of states is related to the structure factor by the sum rule
\begin{equation}
g(\omega)=\frac{1}{\pi}\sum\limits_{\bf q}S({\bf q}, \omega) .
\label{2we}
\end{equation}
According to definition (\ref{45gto}) $S(0,\omega)=0$ since the
position of center of inertia is conserved and $\sum_i u({\bf
r}_i, t)=0$.

\begin{figure}[h!]
     \includegraphics[scale=0.4]{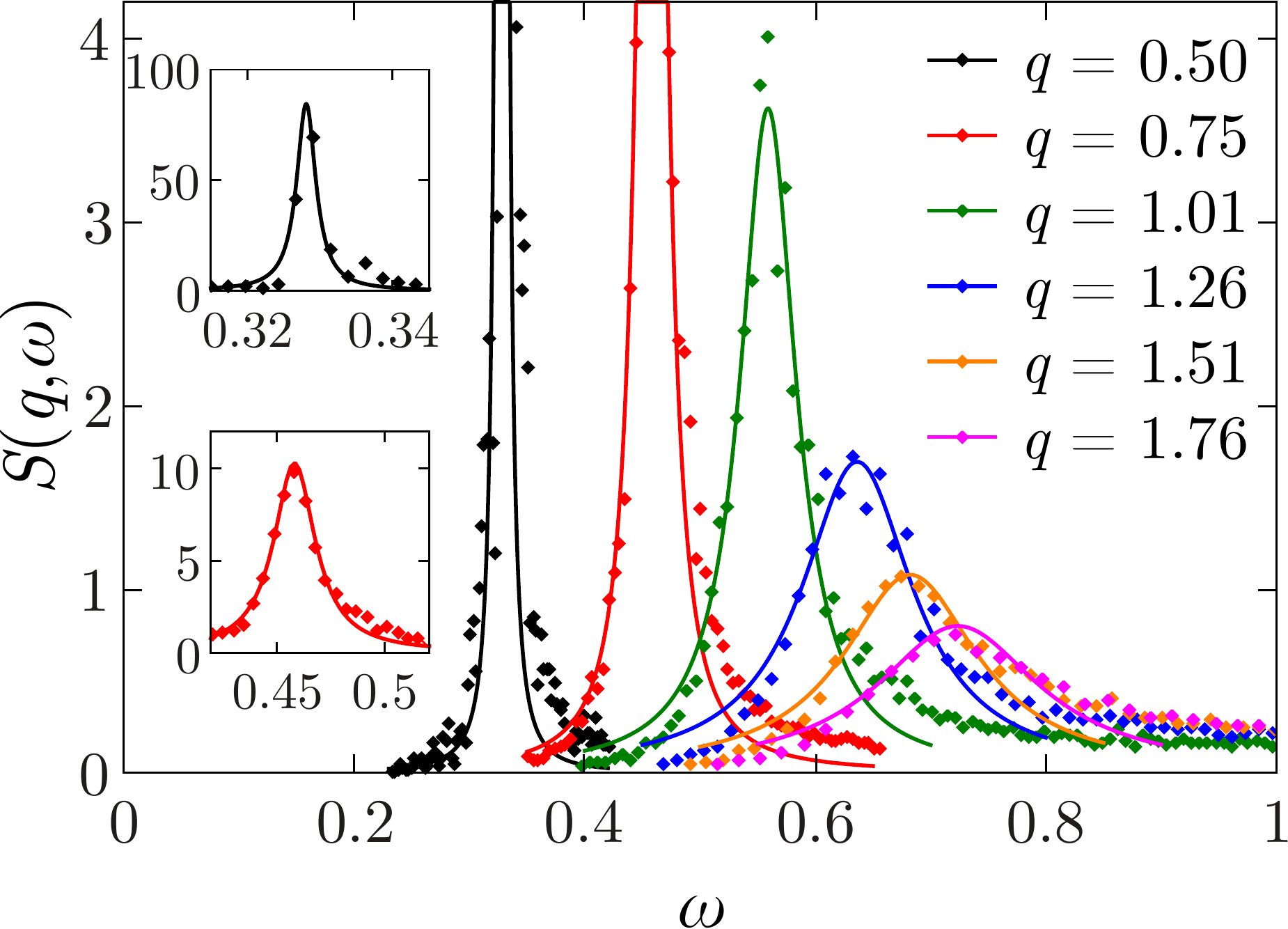}
     \caption{The Lorentz dispersion curves for different wave vectors ${\bf q}\parallel \left<100\right>$
     direction and $\mu=0.1$. Closed diamonds correspond to the calculated values of $S({\bf q}, \omega)$
     and lines are fitting curves according to Eq.~(\ref{s556v}). The number of particles $N=50^3$ (one realization).
     Insets: the Lorentian dispersion curves for $q=0.5$ and $q=0.75$.}
     \label{fig:Lorentz}
\end{figure}

\begin{figure*}[!top!]
    \includegraphics[scale=0.4]{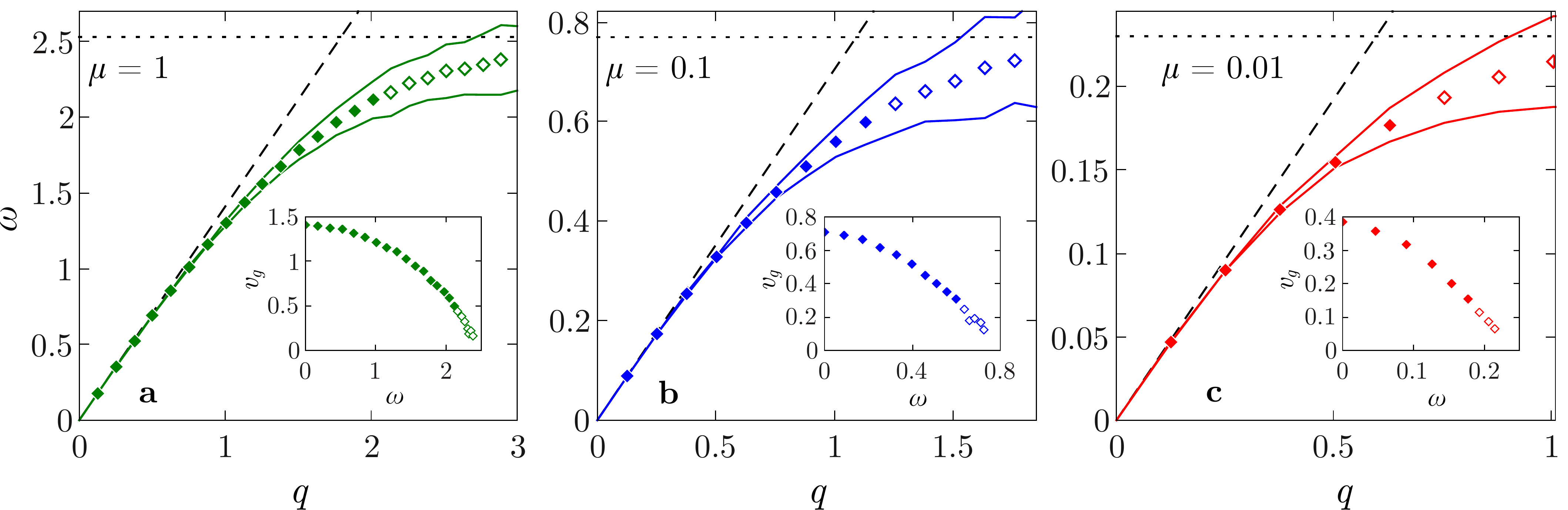}
    \caption{The dependence $\omega_{\bf q}$ on $q$ for ${\bf q}\parallel \left<100\right>$ direction
    for various $\mu$ (1, 0.1, 0.01) in a cubic sample with $N=50^3$ (one realization). Filled and open diamonds
    are the maximums of $S({\bf q},\omega)$ as a function of $\omega$ for each discrete value of $q_n$
    for frequencies below and above the Ioffe-Regel crossover correspondingly (see text below for details).
    Solid lines correspond to halves of the maximums. Dashed lines show $\omega = vq$ linear dependence with sound velocity $v=\sqrt{E}$.
    Horizontal dotted lines correspond to the maximum frequency $\omega_{\rm max}$ in $g(\omega)$ (taken from Fig.~\ref{fig:gwmu}).
    Insets show the group velocity $v_g=d\omega/dq$ as a function of $\omega$.}
    \label{fig:sqw_ph}
\end{figure*}
To analyze phonon excitations, we have found the maximum of
$S({\bf q}, \omega)$ as a function of $\omega$ for each discrete
value of ${\bf q}_n$,  for several values of $\mu$. As an example,
the results for $\mu=0.1$ and  one $\bf q$ direction are shown on
Fig.~\ref{fig:Lorentz}. For the fitting curves we used the Lorentz
distribution
\begin{equation}
S({\bf q}, \omega)\propto \frac{1}{\left(\omega - \omega_{\bf q}
\right)^2 + \left(\Delta \omega \right)^2}. \label{s556v}
\end{equation}

From this fit we can find both the phonon frequency $\omega_{\bf
q}$ and the phonon line width $\Delta\omega$. The results for
$\omega_{\bf q}$ are shown on Fig.~\ref{fig:sqw_ph} for three
values of $\mu$ and ${\bf q}\parallel \left<100\right>$. For
sufficiently small values of wave vector $q$ we see a nice linear
dispersion curve $\omega_q=vq$, with the sound velocity $v$ given
by Eq.~(\ref{s7cvbf}). It is independent of the $\bf q$ direction
(i.e. the sound velocity is isotropic). With increase of $q$, the
frequency $\omega_q$ shows a pronounced negative dispersion of the
group velocity $v_g=d\omega_q/dq$ and approaches the maximum
frequency $\omega_{\rm max}$ where the dependence $\omega_q$
saturates. In this $\bf q$ region we observed a weak anisotropy of
the dispersion curves for $\mu=1$. At smaller values of $\mu$ the
dependence $\omega_{\bf q}$ is isotropic. Since $\omega_{\rm
max}\propto \sqrt{\mu}$, the vertical axis on
Fig.~\ref{fig:sqw_ph} scales approximately as $\sqrt{\mu}$ and the
horizontal axis scales as $\mu^{1/4}$ (sound velocity
$v\propto\sqrt{E}\propto \mu^{1/4}$, and $q_{\rm max}\approx
\omega_{\rm max}/v \propto \mu^{1/4}$ as well).

The strong negative dispersion of the group velocity  $v_g$ for
big $q$ values can be explained by {\em avoided crossing
principle} (or level repulsion effect) due to the coupling of
phonons to quasilocal vibrations near frequency $\omega_{\rm
max}$, corresponding to  sharp maximum in DOS $g(\omega)$ (see
Fig.~\ref{fig:gwmu}). Similar phenomenon exists in polariton
physics~\cite{polariton}. The dip in the participation ratio
$P(\omega)$ for $\mu=0.001$, $\mu=0.01$ and $\mu=0.1$ at
$\omega\approx\omega_{\rm max}$ (see Fig.~\ref{fig:PR}) evidences
in favor of this idea. The vibrations inside the dip correspond to
frequencies near $\omega_{\rm max}$ and have smaller participation
ratio than the others. Therefore they can be referred to as
quasilocal vibrations. In the following we will see that this
strong scattering is also responsible for the deep minimum in the
diffusivity  $D(\omega)$ at $\omega\approx\omega_{\rm max}$ (see
Fig.~\ref{fig:Dw_mu}).

The negative dispersion of the group velosity $v_g$ is responsible
also for the pronounced rise of the phonon DOS above the
$\omega^2$ dependence, given by Eq.~(\ref{st56}). It is clearly
seen on the Fig.~\ref{fig:gwmu}. Indeed, taking the dispersion
into account and disregarding weak anisotropy (taking place only
for $\mu=1$) we can write instead of Eq.~(\ref{st56})
\begin{equation}
g_{\rm ph}(\omega)=
\frac{1}{2\pi^2}\frac{q^2(\omega)}{v_g(\omega)}. \label{d56v}
\end{equation}
Here $v_g(\omega)=d\omega/dq$ is the group velocity shown in
Insets on Fig.~\ref{fig:sqw_ph}. Taking for $q(\omega)$ and
$v_g(\omega)$  the data from Fig.~\ref{fig:sqw_ph} we obtain the
points (filled and open diamonds) shown on Fig.~\ref{fig:gwmu}.
Since they perfectly coincide with numerical data for $g(\omega)$
below $\omega_{\rm max}$, we conclude that {\em all} the
excitations in the phonon gap belong to phonons (with nonlinear
dispersion at higher values of $q$).

\begin{figure}[h!]
     \includegraphics[scale=0.4]{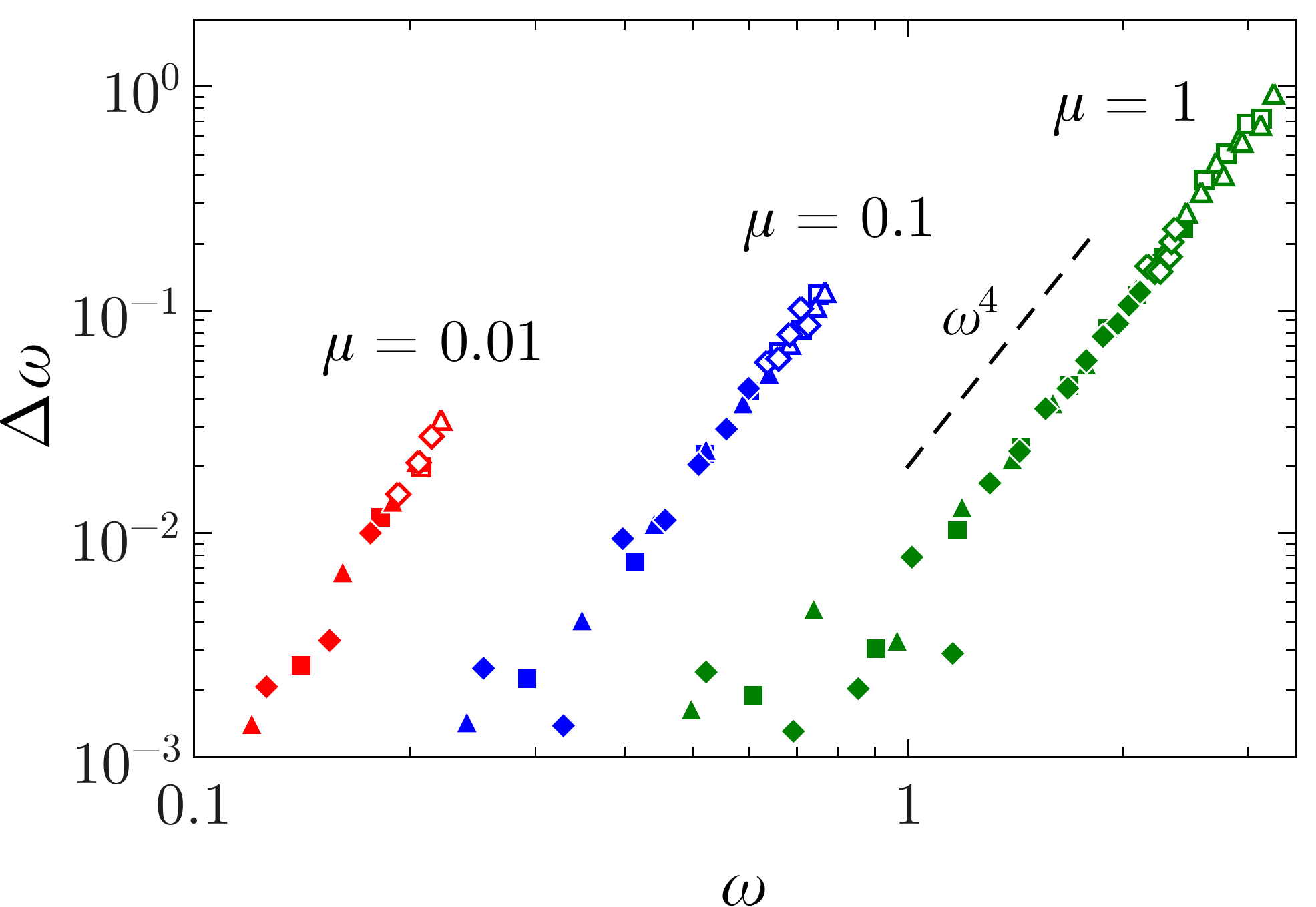}
     \caption{The phonon line width $\Delta\omega$ as a function of $\omega$ for different $\mu$ in cubic sample
     with $N=50^3$ (one realization). Different symbols correspond to different $\bf q$ directions. $\rhomb$ for
     ${\bf q}\parallel \left<100\right>$, $\triangle$ for ${\bf q}\parallel \left<110\right>$, $\square$  for
     ${\bf q}\parallel \left<111\right>$. Filled and open symbols refer to excitations below and above the Ioffe-Regel
     crossover frequency $\omega_{\rm IR}$ correspondingly (see text for details).}
     \label{fig:Dw(w)}
\end{figure}
The phonon line width $\Delta\omega$ can be also found from fits
similar to those shown on Fig.~\ref{fig:Lorentz}. It is related to
the phonon life time $\tau=1/2\Delta\omega$. The factor 2 takes
into account that $\Delta\omega$ corresponds to decay of the
amplitude of  the vibration. The results are shown on
Fig.~\ref{fig:Dw(w)}. As follows from this figure,
$\Delta\omega\propto \omega^4$ and does not depend on the
direction of $\bf q$. We think that this frequency dependence is
not due to Rayleigh scattering of phonons on a static disorder. In
such a  case $\Delta\omega$ would be proportional to $q^4$. Due to
nonlinear dispersion in $\omega_{\bf q}$, these dependencies do
not correspond to each other. More likely, the phonon line width
is due to strong resonant scattering of phonons by quasilocal
vibrations responsible for the sharp peak in the DOS, similar to
those introduced in~\cite{Buchenau1}. The deep minimum in the
diffusivity $D(\omega)$ around frequency $\omega_{\rm max}$ also
supports this idea (see Fig.~\ref{fig:Dw_mu}). We hope to
investigate this important question in future work.

With known value of  $\Delta\omega$, the phonon mean free path
$l(\omega)$ can be calculated as follows
\begin{equation}
l(\omega) = v_g\tau=\frac{v_g}{2\Delta\omega}. \label{we35}
\end{equation}
The phonons are well defined excitations if their mean free path
$l(\omega)$ exceeds the phonon wave length $\lambda=2\pi/q$
(Ioffe-Regel criterium for phonons). As we will see in the next
Section, phonons transform to diffusons when $l(\omega)\approx
\lambda/2$. We will call the corresponding crossover frequency as
$\omega_{\rm IR}$.  Fig.~\ref{fig:IR} shows the ratio
$l(\omega)/\lambda$ as a function of $\omega$ for several values
of $\mu$ and different directions of the wave vector $\bf q$. The
boundary between filled and open symbols (the full horizontal
line) corresponds to frequency $\omega_{\rm IR}$. Thus filled and
open symbols on Figs.~\ref{fig:gwmu}, \ref{fig:sqw_ph},
\ref{fig:Dw(w)}, \ref{fig:IR}  belong to phonons with frequencies
below and above the Ioffe-Regel crossover frequency
correspondingly.
\begin{figure}[h!]
     \includegraphics[scale=0.4]{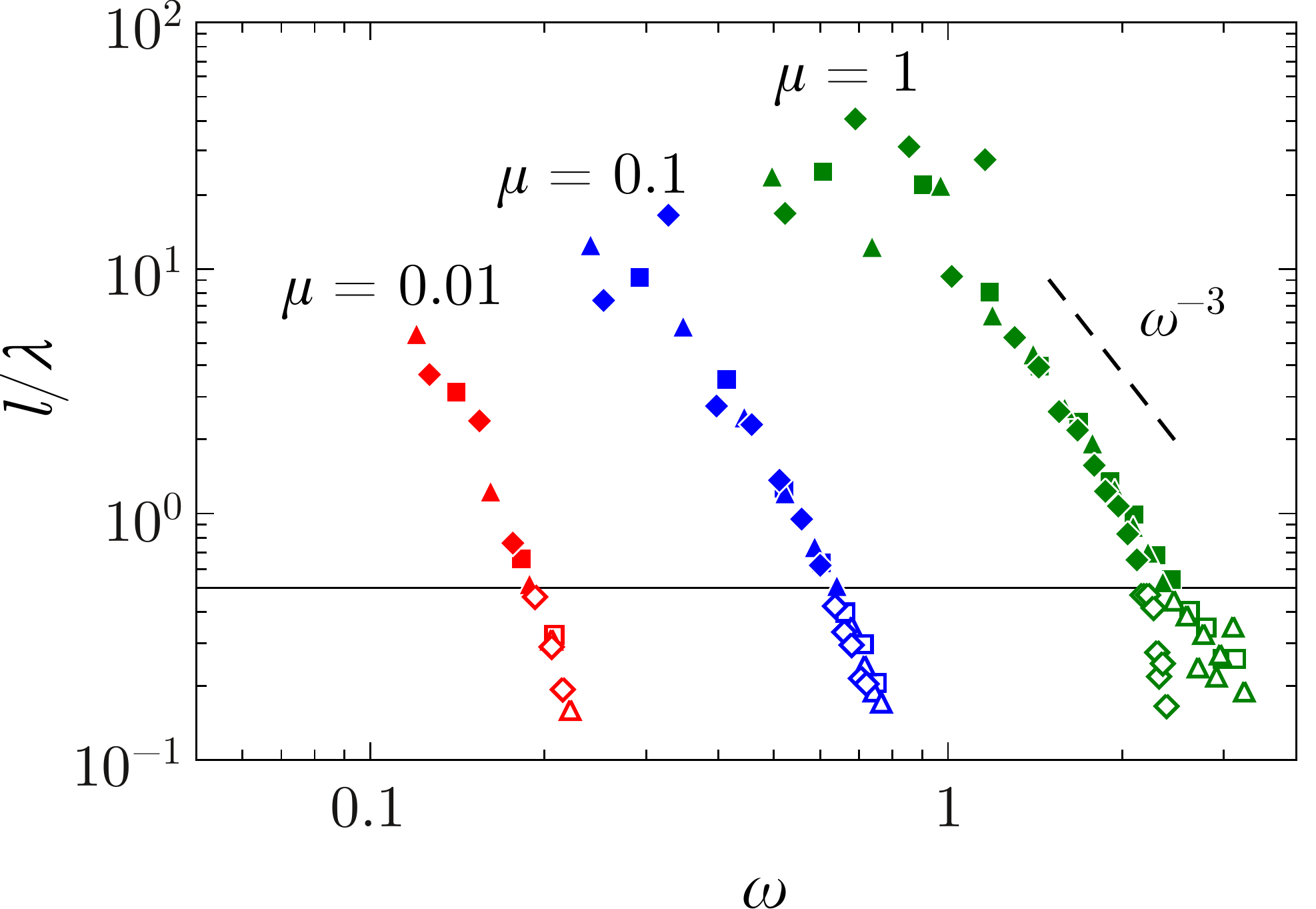}
     \caption{The ratio $l(\omega)/\lambda$ as a function of $\omega$ for different $\mu$. Different symbols
     correspond to different $\bf q$ directions as explained on Fig.~\ref{fig:Dw(w)}. The full horizontal line
     (separating filled and open symbols) corresponds to Ioffe-Regel crossover $l(\omega)=\lambda/2$.}
     \label{fig:IR}
\end{figure}

\begin{figure*}[!top!]
    \includegraphics[scale=0.4]{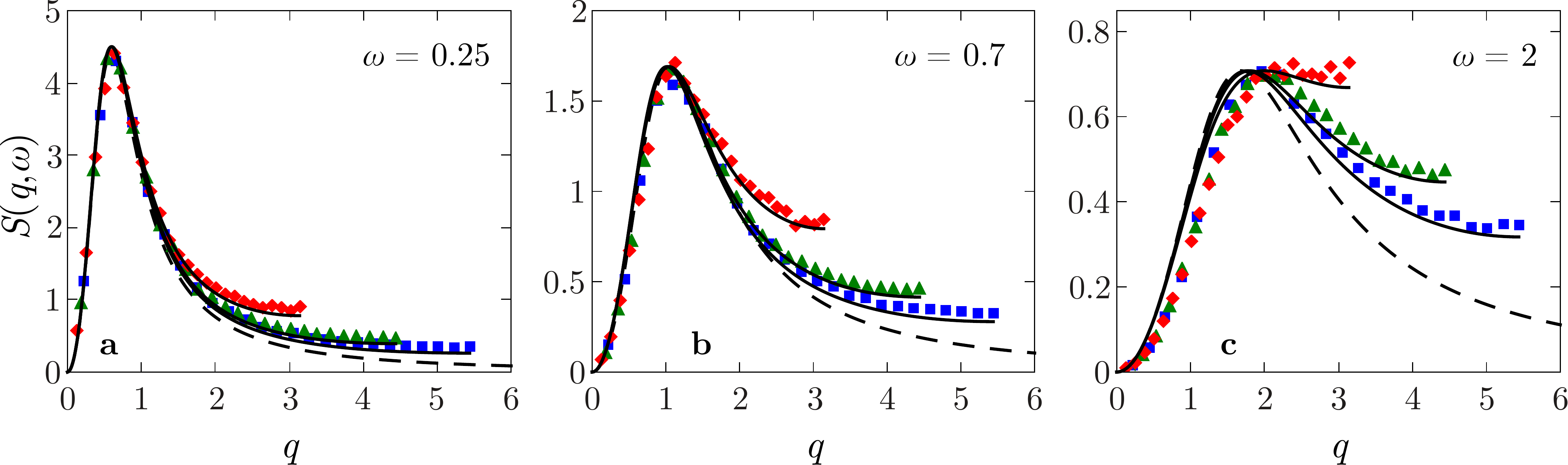}
    \caption{The displacement structure factor $S({\bf q},\omega)$,  Eq.~(\ref{45gto}) (symbols) for $\mu=0$ and for three
    different frequencies. The sample size is $N=50^3$. The averaging is performed over 300 realizations.
    Different symbols correspond to different $\bf q$ directions. $\rhomb$ for
     ${\bf q}\parallel \left<100\right>$, $\triangle$ for ${\bf q}\parallel \left<110\right>$, $\square$  for
     ${\bf q}\parallel \left<111\right>$.  Full lines correspond to the structure factor $S_{\rm rw}({\bf q}, \omega)$
     of the random walk on the lattice given by Eq.~(\ref{s6as}) with $D_{\rm rw}=0.7$. Dashed line corresponds
     to the limit $q\ll 1$ (see Eq.~(\ref{uhg7})).}
    \label{fig:Sqw_RW}
\end{figure*}
Usually in glasses the Ioffe-Regel crossover frequency
$\omega_{\rm IR}$ is correlated with position of the boson peak
$\omega_b$, see~\cite{laermans1, laermans2, laermans3, laermans4,
tanaka} and references therein. It is the frequency where the
reduced DOS $g(\omega)/\omega^2$ has a maximum. We also have a
rather sharp boson peak in our disordered lattices~\cite{our2}.
As follows from Fig.~\ref{fig:gwmu} the left side of the boson
peak is built from phonons having negative dispersion of the group
velocity $d\omega_{\bf q}/d{\bf q}$. Similar conclusion was made
recently for $2d$ and $3d$ Lennard-Jones glasses~\cite{tanguy,
tan2005, monaco}. The right side of the boson peak consists from
diffuson modes shifted from the region of small frequencies
$0<\omega<\omega_{\rm max}$ by additional term $\mu M_0$ and further
modified  by interaction with phonons. But more work is necessary to
elucidate the precise structure of these modes.

The frequencies $\omega_{\rm max}$, $\omega_{\rm IR}$, and
$\omega_b$ are collected in Table~\ref{tab1} for different $\mu$.
As we can see from the table, $\omega_{\rm IR}$ is close to the
frequency $\omega_{\rm max}$ and to the position of the boson peak
$\omega_b$. Above $\omega_{\rm IR}$ phonons cease to exist  as
well defined excitations. They are smoothly transformed to
diffusons which we will consider in the next Section. The relative
number of phonons in the lattice can be estimated as follows
\begin{equation}
N_{\rm ph} = \int\limits_{0}^{\omega_{\rm IR}}g(\omega) d\omega .
\label{pnononsnumber}
\end{equation}
These values are also given in the Table~\ref{tab1}. We see that
for all investigated values of $\mu$ the relative number of
phonons in the lattice is small. It is in agreement with similar
estimates for amorphous silicon~\cite{Nature5}.
\begin{table}[h!]
    \begin{tabular}{|*{5}{@{\hspace{0.4cm}}c@{\hspace{0.4cm}}|}}
        \hline
        $\mu$ & $\omega_{\rm max}$ & $\omega_b$ & $\omega_{\rm IR}$ & $N_{\rm ph}$ \\
        \hline
        1     & 2.5  & 2.4  & 2.2$^*$ & 0.12 \\
        0.1   & 0.78  & 0.74  & 0.62 & 0.027 \\
        0.01  & 0.23  & 0.23  & 0.19 & 0.0066 \\
        0.001 & 0.072 & 0.07 &   &   \\
        \hline
    \end{tabular}
        \caption{The frequency of maximum in DOS $\omega_{\rm max}$, the frequency of the Ioffe-Regel
        crossover $\omega_{\rm IR}$ and the boson peak frequency $\omega_b$ for various $\mu$. Star $^*$
        means that $\omega_{\rm IR}$ was found for ${\bf q}\parallel\left<100\right>$ direction. $N_{\rm ph}$ is
        a relative number of phonons in the lattice.}
        \label{tab1}
\end{table}

\section{Diffusons}
\label{diff} In this section we are going to consider properties
of diffusons. As is well known, the diffusion phenomenon usually
takes place for physical quantities which are conserved. In a free
closed mechanical system we have two integrals of motion, momentum
and energy. Therefore one should discriminate between diffusion of
momentum and energy.

\subsection{Diffusion of momentum}
\label{diffmomentum}

First let us consider diffusion of momentum. Usually the diffusion
of momentum is related to viscosity in the system. When all
particle masses being equal ($m_i=1$), the diffusion of momentum
is equivalent to the diffusion of particle displacements.  It is
because in our system the position of the center of inertia is
conserved and we can put it at the origin of the coordinate
system. Then the sum of all particle displacements vanishes
\begin{equation}
\sum\limits_i u_i(t)=0, \label{xty5}
\end{equation}
i.e. it is an integral of motion. The diffusion of displacements
in this case looks like a diffusion of ''particles'' in a lattice
where the total number of particles is conserved.

By analogy with diffusion of ''particles'' the information about
diffusivity of displacements is absorbed in the displacement
structure factor $S({\bf q}, \omega)$ (\ref{45gto}). We remind
that to calculate this structure factor we ascribed at the initial
moment $t=0$ the random displacements to all the particles with
Gaussian distribution (with zero mean  and unit variance) and
velocities equal to zero. So the condition (\ref{xty5}) at $t=0$
was satisfied. Therefore let us analyze now this structure factor
in the diffuson frequency range.

Consider first the case of $\mu=0$ when phonons are absent and
only diffusons are present in the lattice. Fig.~\ref{fig:Sqw_RW}
shows the structure factor $S({\bf q}, \omega)$ as a function of
wave vector $q$ for three different directions in ${\bf q}$ space
(symbols) and for three different frequencies $\omega$. Let us
compare this displacement structure factor with structure factor
of the random walk $S_{\rm rw}({\bf q}, \omega)$ on the lattice.

As was shown in~\cite{difcondmat}  for the case of the random walk
on a lattice, $S_{\rm rw}({\bf q}, \omega)$ is given by expression
\begin{equation}
S_{\rm rw}({\bf q}, \omega)= \frac{2\Gamma({\bf q})}{\omega^2 +
\Gamma^2({\bf q})}. \label{s6as}
\end{equation}
It is a Lorentzian, with a width $\Gamma({\bf q})$ given by
\begin{equation}
\Gamma({\bf q})=D_{\rm rw}Q^2({\bf q}) , \label{s7gh}
\end{equation}
where $D_{\rm rw}$ is a diffusion constant of the random walk. In
a simple cubic lattice (with lattice constant $a_0=1$) the
function $Q({\bf q})$ reads
\begin{equation}
Q({\bf q})=2\sqrt{\sin^2\frac{q_x}{2} + \sin^2\frac{q_y}{2} +
\sin^2\frac{q_z}{2}}. \label{tu67}
\end{equation}
For small values of $q\ll 1$, $Q({\bf q})=q$ and in the continuum
limit we have the well known result for the diffusion structure
factor
\begin{equation}
S_{\rm rw}({\bf q}, \omega)= \frac{2D_{\rm rw}q^2}{D^2_{\rm
rw}q^4+\omega^2}. \label{uhg7}
\end{equation}
Let us note that the structure factor (\ref{s6as}) has a maximum
at ${\bf q}$ values obeying the condition
\begin{equation}
\omega=\Gamma({\bf q})=D_{\rm rw}Q^2({\bf q}). \label{2wdr}
\end{equation}
We can specify it as a {\em dispersion law for diffusons}. The
width of the maximum is $\Gamma({\bf q})$. For $q\ll 1$,
$\Gamma({\bf q})= D_{\rm rw} q^2$.

\begin{figure}[!h!]
    \includegraphics[scale=0.4]{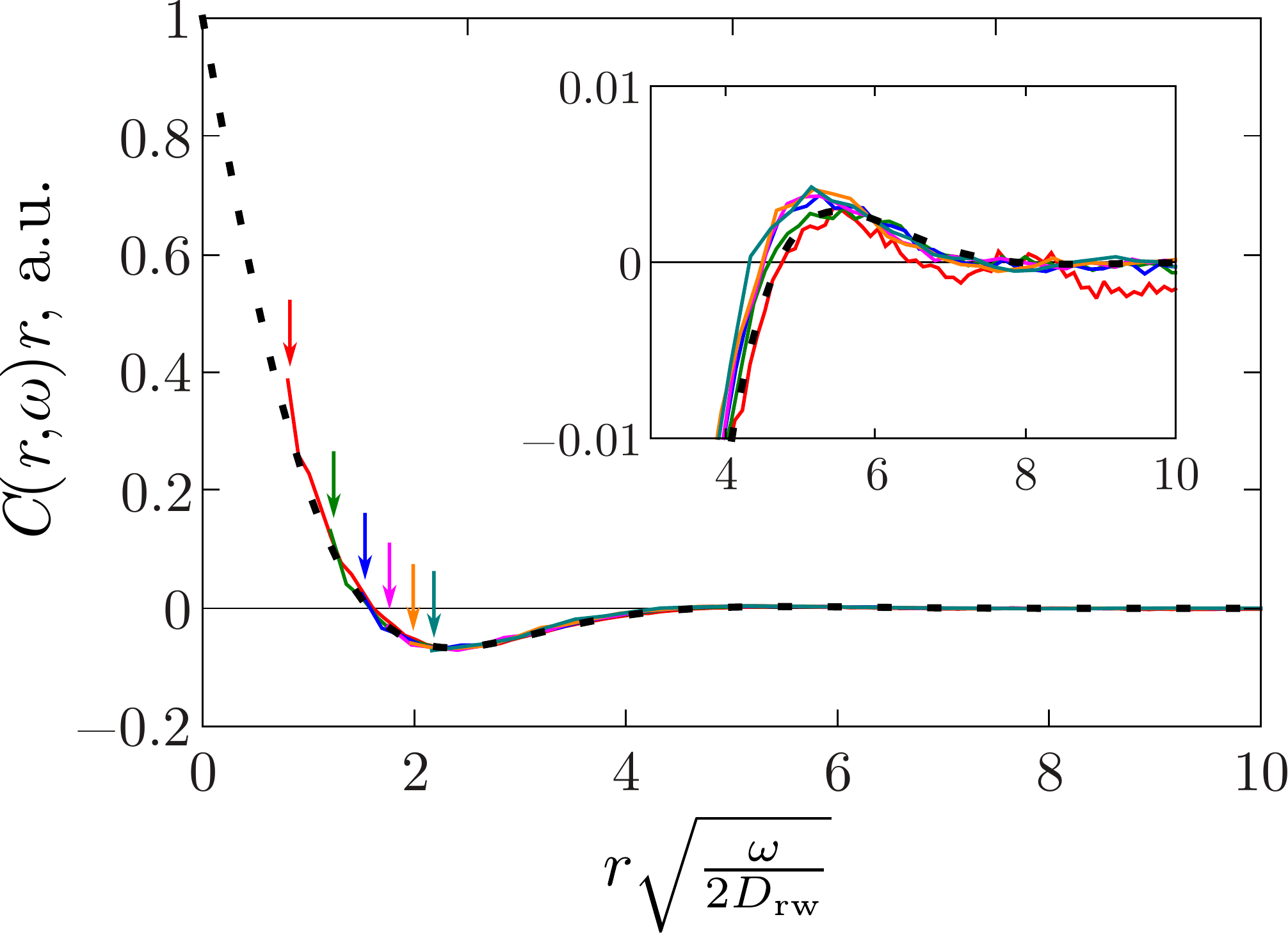}
    \caption{The correlation function $C({\bf r},\omega)$ for $\mu=0$ and six frequencies $\omega$
    (0.14, 0.31, 0.49, 0.66, 0.84, 1.01) for sample with $N=50^3$ particles averaged over 300 realizations.
    The full lines are our numerical results obtained from Eq.~(\ref{45gto}). Each line starts from $r=r_{\rm min}$
    which is about 2.5 interatomic distances (marked by arrows). The dashed line corresponds to Eq.~(\ref{eq:Corr}) with $D_{\rm rw} = 0.7$.}
    \label{fig:sqwCorr}
\end{figure}

\begin{figure*}[!top!]
    \includegraphics[scale=0.4]{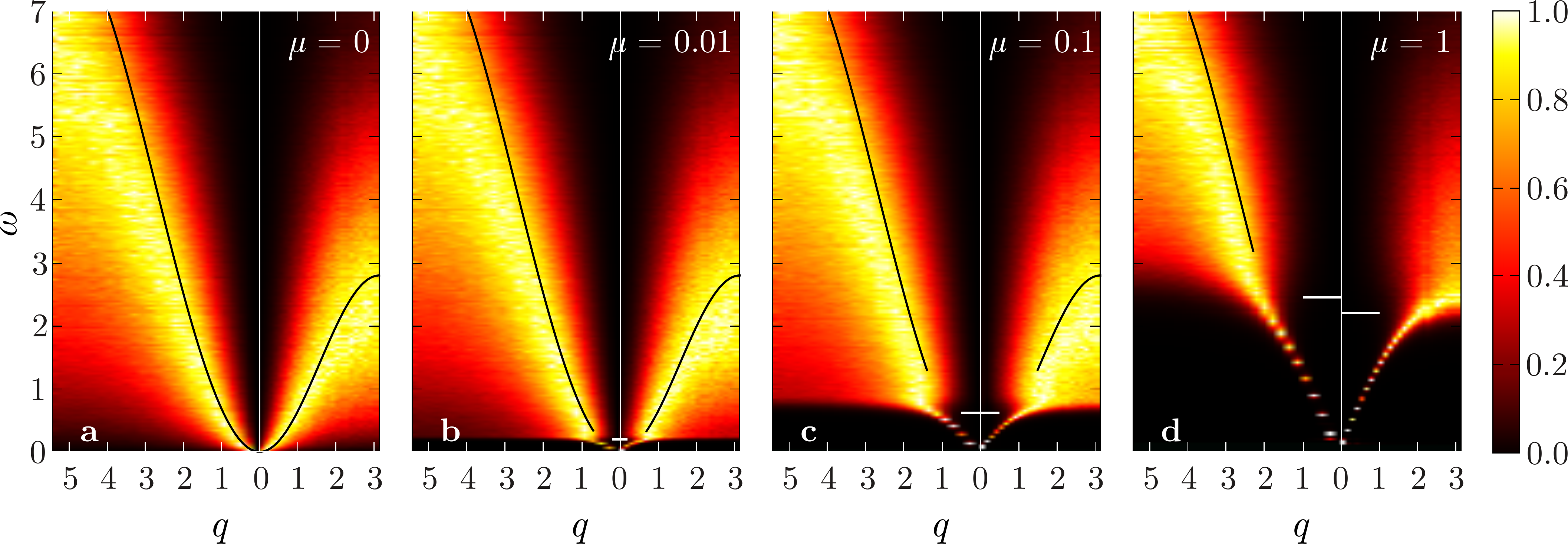}
    \caption{The normalized structure factor $S_n({\bf q},\omega)$ as a function of $q$ for some direction in $\bf q$
    space and for each frequency $\omega$ for various values of $\mu$ (0, 0.01, 0.1, 1). The  sample size is $N=50^3$. The averaging
    is performed over 100 realizations.
     Left sides of all plots are for ${\bf q}\parallel \left<111\right>$, right sides are for ${\bf q}\parallel \left<100\right>$.
     White horizontal dashes show the Ioffe-Regel crossover frequency $\omega_{\rm IR}$. For $\mu=1$ the frequency $\omega_{\rm IR}$
     is slightly different for different ${\bf q}$ directions. Black full line corresponds to Eq.~(\ref{2wdr}) for the random walk on
     a simple cubic lattice with diffusion constant $D_{\rm rw}=0.7$.}
    \label{fig:sqw}
\end{figure*}

A comparison of the displacement structure factor $S({\bf q},
\omega)$, (\ref{45gto}), and the structure factor of the random
walk $S_{\rm rw}({\bf q}, \omega)$, (\ref{s6as}), is shown on
Fig.~\ref{fig:Sqw_RW}. One fitting parameter was the diffusion
coefficient $D_{\rm rw}$ in Eq.~(\ref{s7gh}). From comparison of
these data we  obtained $D_{\rm rw}\approx 0.7$. It means that the
diffusion coefficient of particle displacements $D_u\approx 0.7$
(see Section~\ref{disc}). Another fitting parameter was a height
$h(\omega)$ of the random walk structure factor in the maximum.
According to Eq.~(\ref{s6as}), in the maximum $\Gamma({\bf
q})=\omega$ and $h(\omega)=1/\omega$, but to fit the data points
on  Fig.~\ref{fig:Sqw_RW} we used slightly higher values of
$h(\omega)$.

The small difference between $h(\omega)$ and $1/\omega$ can be
explained by different frequency dependencies of the density of
states $g(\omega)$ for vibrations and for the random walk
(following from the sum rule similar to Eq.~(\ref{2we})). As we
can see from the figure, for the investigated frequencies the fit
is perfect. With increasing frequency above $\omega\approx 2-3$,
the fitting becomes more and more poor since we approach the
localization threshold at $\omega_{\rm loc}\approx 5.5\pm 0.5$
(see below) which is not described well by a simple model of
Markovian random walk.

Now let us consider a behavior of a correlation function. The
correlation function of particle displacements at some frequency
$\omega$, expressed through eigenvectors $e_{\bf r}(\omega)$ of
the dynamical matrix $M$, reads
\begin{equation}
        C({\bf r},\omega) = \sum\limits_{{\bf r}'}e_{{\bf r}'+{\bf r}}(\omega)e_{{\bf r}'}(\omega).
\end{equation}
It is a Fourier transform of the displacement structure factor
(\ref{45gto})
\begin{equation}
    C({\bf r},\omega) = \frac{1}{8\pi^4}\int S({\bf q},\omega) e^{i {\bf qr}} d{\bf q}.
    \label{s55fg}
\end{equation}

Let us compare this correlation function with correlation function
of the random walk. For distances bigger than the period of the
lattice ($a_0=1$) we can make use of the limit of small $q\ll 1$
and integrate   Eq.~(\ref{uhg7}) for the random walk structure
factor taken in approximation of continuous media. As a result, we
derive
\begin{equation}
    C_{\rm rw}({\bf r},\omega) = \frac{\exp\left(-r\sqrt{ \frac{\displaystyle\omega}{\displaystyle2D_{\rm rw}}}\,\right)\cos\left(r\sqrt{\frac{\displaystyle\omega}{\displaystyle2D_{\rm rw}}}\,\right)}{2\pi^2 r D_{\rm rw}}. \label{eq:Corr}
\end{equation}

Fig.~\ref{fig:sqwCorr} shows a good agreement of our correlation
function (\ref{s55fg}) with the correlation function of the random
walk (\ref{eq:Corr}). For all investigated frequencies the
numerical data collapse together and become indistinguishable from
the theoretical prediction (\ref{eq:Corr}).  We can see also on
this figure the anticorrelation phenomenon (the region of negative
values of the correlation function). As follows from
Eq.~(\ref{eq:Corr}), the correlation function of the random walk
changes its sign for the first time at
\begin{equation}
r\sqrt{\frac{\displaystyle\omega}{\displaystyle2D_{\rm
rw}}}=\frac{\pi}{2} . \label{s5tt}
\end{equation}
It is also in a good agreement with our numerical results.
Therefore we can call a corresponding value of $r$ found from
Eq.~(\ref{s5tt}) as a {\em radius of diffuson}. It is a typical
size of the regions vibrating with frequency $\omega$ and having
the same sign of all particle displacements. According to
(\ref{s5tt}), the radius of diffuson is given by
\begin{equation}
r_{\rm d}(\omega)=\frac{\pi}{\sqrt{2}}\sqrt{\frac{D_{\rm
rw}}{\omega}}\propto \omega^{-1/2} . \label{q8bn}
\end{equation}
At $\omega=0$ the correlation function (\ref{eq:Corr}) decays
slowly as $1/r$. In disordered systems at critical point the
correlation function decays as $C(r)\propto 1/r^{d-D_2}$ where $d$
is the space dimension and $D_2$ is a correlation dimension. From
this we conclude that in our case $D_2=2$ what corresponds to
diffusion.

Now let us analyze the displacement structure factor $S({\bf q},
\omega)$ for $\mu\ne 0$. For better visual effect we will show a
map of the function  $S({\bf q}, \omega)$ on the plane ($\omega$,
$q$) for different directions in $\bf q$ space. To do that, for
each frequency $\omega$ we have found the maximum $S({\bf q},
\omega)$ as a function of $q$ along some directions in  $\bf q$
space. Then we normalized function $S({\bf q}, \omega)$ along this
line $\omega$=const to the magnitude of this maximum.

The results are shown on Fig.~\ref{fig:sqw} for four different
values of $\mu$ and two directions in $\bf q$ space. The white
color corresponds to the maximum when normalized structure factor
$S_n({\bf q}, \omega)=1$ while the black color to the case where
$S_n({\bf q}, \omega)=0$. For $\mu\ne 0$ we can see clearly two
types of excitations in the lattice. At low enough frequencies,
below $\omega_{\rm IR}$, we see phonons with well defined
dispersion law $\omega_{\bf q}$, the same as in the previous
Section. At the Ioffe-Regel crossover frequency $\omega_{\rm IR}$,
the structure factor strongly broadens and  phonon dispersion line
disappears. Above $\omega_{\rm IR}$ the displacement structure
factor coincides well with the structure factor for $\mu=0$ case
shown on Fig.~\ref{fig:sqw}a, which corresponds to diffusons. The
maximum of the normalized structure factor $S_n({\bf q},\omega)$
(white regions) agrees well with Eq.~(\ref{2wdr}) (with the same
diffusion coefficient $D_{\rm rw}$) giving the maximum of the
random walk structure factor $S_{\rm rw}({\bf q}, \omega)$ (black
line). It means that diffusion coefficient of particle
displacements is independent of $\mu$. Deviations from $S_{\rm
rw}({\bf q}, \omega)$ take place at high frequencies near the
localization threshold.

For $\mu\ne 0$ the radius of diffuson (\ref{q8bn}) takes a maximum
value at $\omega\approx\omega_{\rm IR}$. At smaller frequencies we
have well defined phonons. Since $\omega_{\rm
IR}\propto\sqrt{\mu}$ and $D_{\rm rw}\approx 1$ we can write for
$0<\mu\lesssim 1 $
\begin{equation}
r_{\rm d}(\omega_{\rm IR})\equiv r_c \simeq \sqrt{D_{\rm
rw}/\omega_{\rm IR}}\simeq \mu^{-1/4}. \label{3b6g}
\end{equation}
The value $r_c$ plays a role of correlation length in our lattice.
It diverges when $\mu\to 0$. The physical meaning of this length
is that it by the order of the value coincides with the
Ioffe-Regel wave length $\lambda_{\rm IR}=2\pi/q_{\rm IR}$
corresponding to frequency $\omega_{\rm IR}$ (see
Section~\ref{scaling}). Samples with size smaller than $r_c$ have
no phonon-like modes at al.

\begin{figure}[h!]
    \includegraphics[scale=0.4]{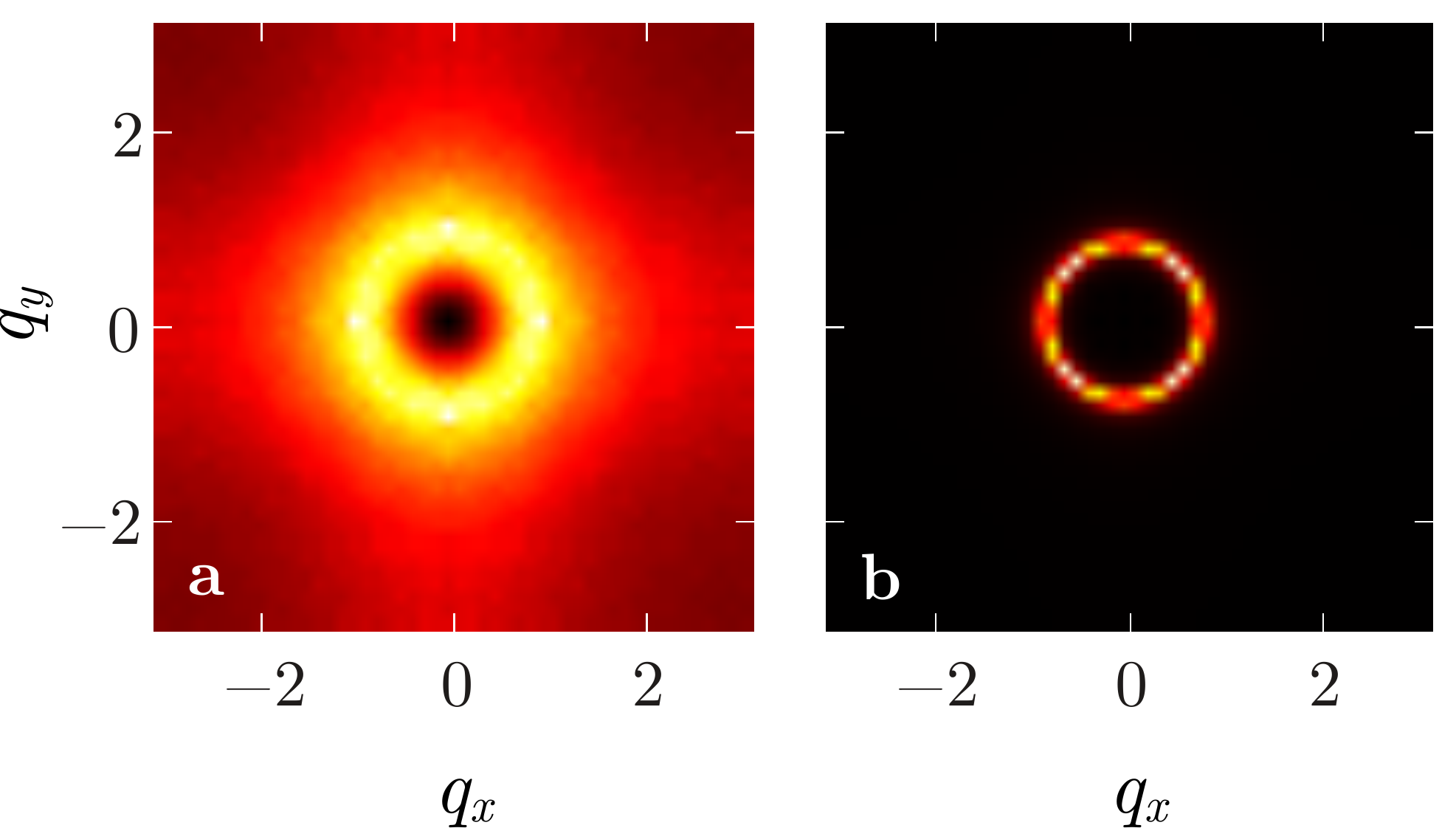}
    \caption{The same normalized structure factor $S_n({\bf q},\omega)$ as on Fig.~\ref{fig:sqw}
    but in $\bf q$ space in plane $q_xq_y$ ($q_z=0$) for $\omega=0.5$.  The left picture corresponds to $\mu=0$ (a) and the right to $\mu=0.1$ (b).}
    \label{fig:qxqy}
\end{figure}

To compare phonon and diffuson structure factors, a cross section
of the structure factor $S_n({\bf q},\omega)$ in $\bf q$ space for
$q_z=0$ and frequency $\omega=0.5$ is shown on Fig.~\ref{fig:qxqy}
for $\mu=0$ and $\mu=0.1$. At the left side (a) of this figure we
see the structure factor of diffuson. On the right side we see the
structure factor of phonon (b). As compared with phonon structure
factor, the diffuson structure factor is much more broadened.

\subsection{Diffusion of energy}
\label{ub55p}

Now let us consider the diffusion of energy. The diffusion of
energy is different from diffusion of particle displacements (see
Section~\ref{disc}). The first approach to calculate the
diffusivity of energy $D(\omega)$ for vibrations with frequency
$\omega$ is a direct numerical solution of Newton's equations.
For that we have used the Runge-Kutta-4 method with time step
$\Delta t=0.01$ applied to a cubic sample with $N=L\times L\times
L$ particles (lattice constant $a_0=1$) and with free boundary
conditions along the $x$ direction. Along other two directions we
take the  periodic boundary conditions.

\begin{figure}[h!]
    \includegraphics[scale=0.4]{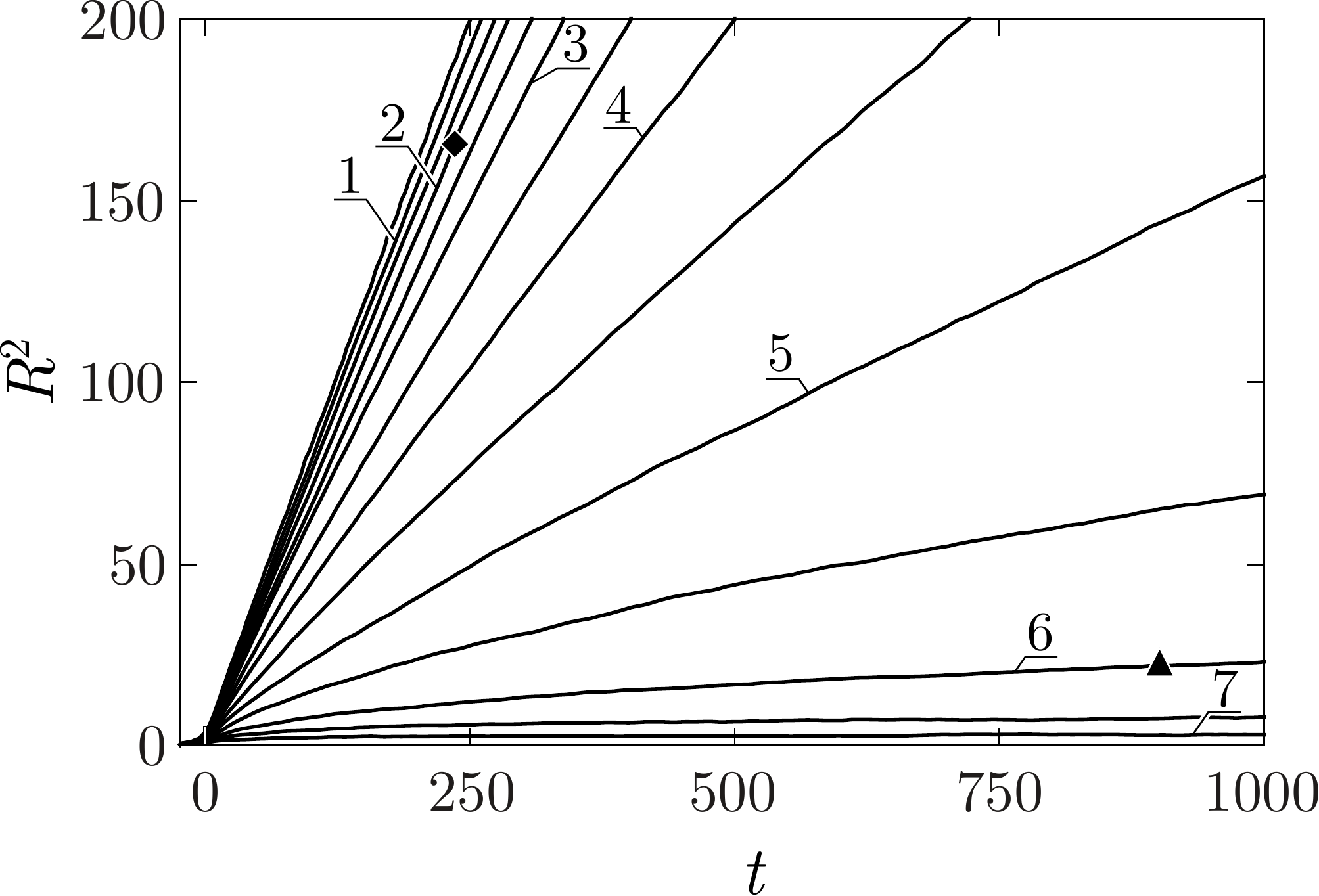}
    \caption{The dependence of $R^2(t)$ in the case of $\mu = 0$ for one sample with $N=100\times 100\times 100$ particles and $14$
    different frequencies $\omega=0.5, 1, 1.5, \ldots, 7$ (from top to bottom). The numbers indicate integer frequencies. The slope of each
    line corresponds to each black dot in Fig.~\ref{fig:Dw}. Two points at $\omega=2$ and $\omega=6$ correspond to two distributions of energy
    $E(x,t)$ over the sample for delocalized and localized modes correspondingly. They are shown on Fig.~\ref{fig:Ex} (see below).}
    \label{fig:fan}
\end{figure}

Assuming zero initial conditions for displacements and velocities
of all the particles, let us apply external forces with frequency
$\omega$ and random phases $\varphi_i$ to all the particles in the
central layer $x=0$ of our sample~\cite{central}
\begin{equation}
    f^{\rm ext}_i(t)=\sin(\omega t+\varphi_i)\exp\left(-\frac{t^2}{2T^2}\right) \label{eq:Nf}
\end{equation}
where $\omega T\gg 1$. The right and the left sides of the sample
have coordinates $x_{\rm r,l}=\pm L/2$. In such a way we excite
vibrations with frequencies near frequency $\omega$ distributed in
a small frequency interval $(\omega-1/T,\, \omega+1/T)$. In
calculations we used $T=5$ for all frequencies $\omega$. We
started our calculations at time $t_{0}=-5T$ when the external
force is still negligible.

After applying the force to the central layer $x=0$, vibrations
will spread to the left and to the right ends of the sample.  The
average squared distance to the energy diffusion front we define
as usual
\begin{equation}
    R^2(t)=\frac{1}{E_{\rm tot}}\sum\limits_{i=1}^N x_i^2 E_i(t) = \frac{1}{E_{\rm tot}} \int\limits_{-L/2}^{L/2}x^2 E(x,t)dx .
    \label{eq:NR2}
\end{equation}
Here $x_i$ is the $x$ coordinate of the $i$-th particle, $E_i(t)$
is the energy of $i$-th particle and sum is taken over all
particles in the sample. $E_{\rm tot}=\sum_i E_i(t)$ is the total
energy of the system. It is independent of time after the external
force $f^{\rm ext}_i(t)$ becomes negligibly small (i.e. for
$t>5T$).

The energy of $i$-th particle $E_i(t)$  we define as a sum of the
kinetic energy and a half of the potential energy of connected
bonds ($m_i=1$)
\begin{equation}
    E_i(t) = \frac{v_i(t)^2}{2}-\frac{1}{4}\sum\limits_j M_{ij}\big(u_i(t)-u_j(t)\big)^2.
\end{equation}
Here $v_i(t)={\dot u}_i(t)$ is  a particle velocity.  Summation
over all particles in Eq.~(\ref{eq:NR2}) we can divide in two
steps. First we sum over all particles in the layer $x$ and then
we sum over all layers. Let  $E(x,t)$ be a total energy confined
to the layer $x$ at time $t$. Having in mind that in our case we
have lattice constant $a_0=1$ and sample size $L\gg 1$, we can
change summation over different layers to integration over
coordinate $x$ for times where $R(t)\gg 1$.
\begin{figure}[h!]
    \includegraphics[scale=0.4]{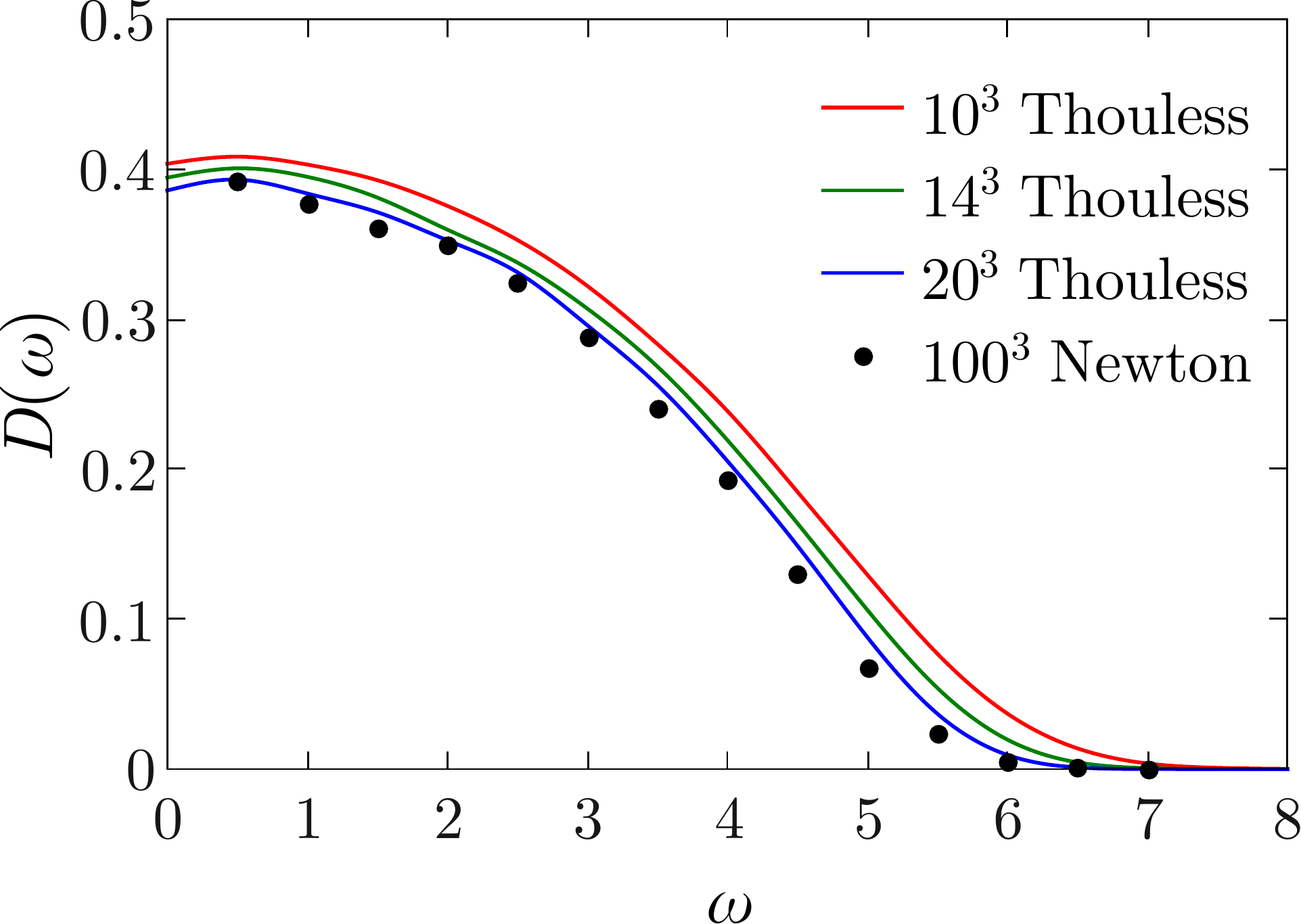}
    \caption{The dependence of diffusivity $D(\omega)$ on $\omega$ for $\mu = 0$. Black dots are calculated by the
    direct solution of Newton's equations from Eqs.~(\ref{eq:NR2}, \ref{eq:ND}) and Fig.~\ref{fig:fan} for $N=100^3$ particles (one realization).
    Full lines for $N=10^3, 14^3, 20^3$ are calculated using formula of Edwards and Thouless (\ref{eq:ETp}) with $c=1$ (see below).
    Averaging for lines is performed over frequencies in the small interval $(\omega-\delta\omega,\omega+\delta\omega)$ with $\delta\omega=0.25$
    and over several thousands realizations.}
    \label{fig:Dw}
\end{figure}

We will apply this method to the case of $\mu=0$ (i.e. for the
lattice without phonons). The results are shown on
Fig.~\ref{fig:fan}. As we can see from the figure for small and
middle frequencies, $R^2(t)\propto t$. Therefore for these
frequencies vibrations  indeed spread  along the $x$ axis by means
of diffusion. The slope of the lines decreases with frequency
$\omega$. For calculating the slope, we take the time interval
$\Delta t$ where, on the one hand $t > 5T$, and on the other hand,
$R \ll L/2$.

From the slope of $R^2(t)$ we can calculate the diffusivity  of
modes $D(\omega)$ using one dimensional formula
\begin{equation}
    R^2(t) = 2D(\omega)t.
    \label{eq:ND}
\end{equation}
This diffusivity is shown by black dots on Fig.~\ref{fig:Dw}. At
small frequencies it is approximately constant, then it decreases
with frequency approaching zero at the localization threshold,
$\omega_{\rm loc}\approx 5.5\pm 0.5$. At higher frequencies above
$\omega_{\rm loc}$ the dependence $R^2(t)$ saturates with
increasing $t$. This indicates localization of the vibrational
modes.

\begin{figure}[h!]
    \includegraphics[scale=0.4]{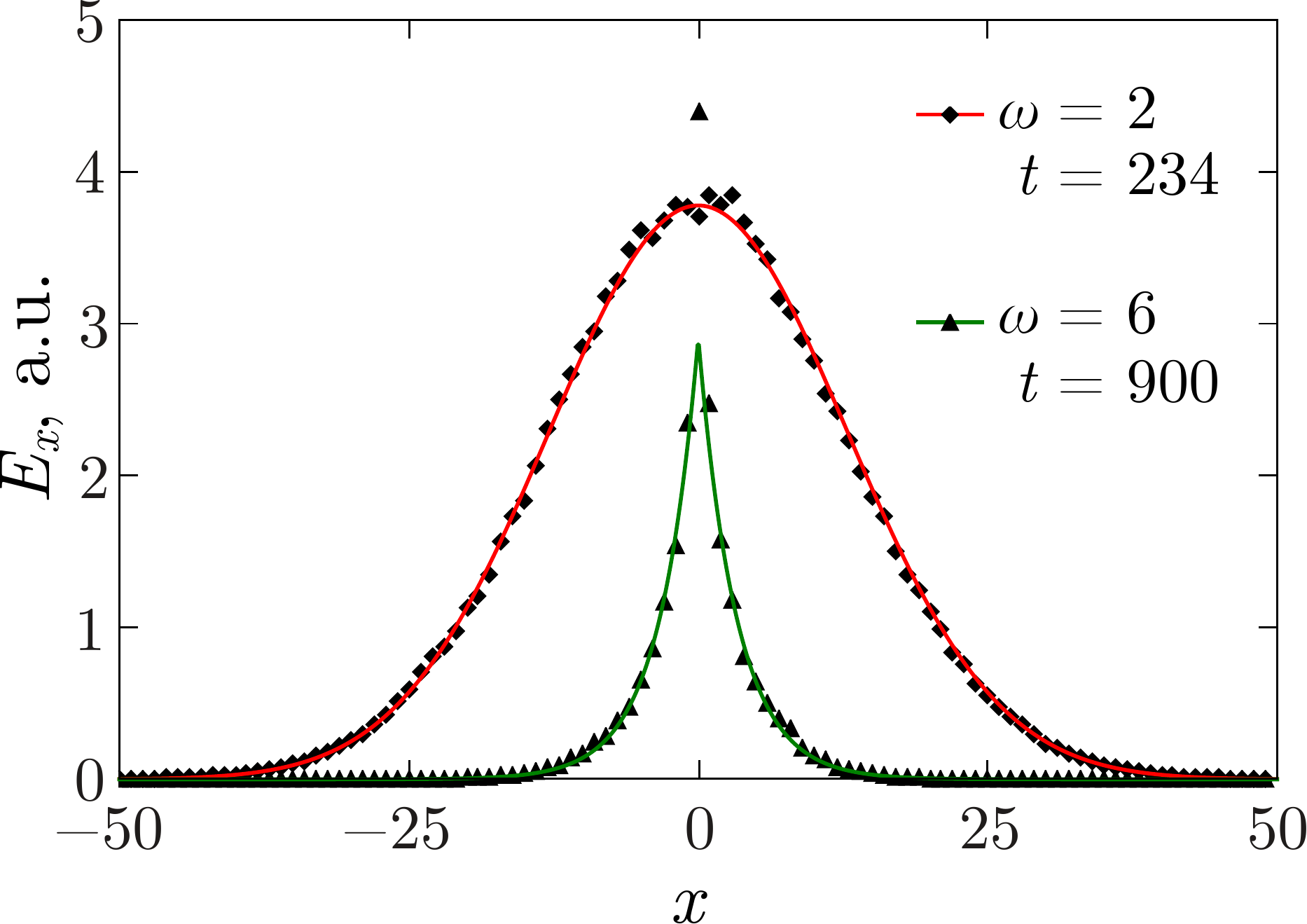}
    \caption{Black points (diamonds and triangles) show the distribution of energy $E(x,t)$ contained in the layer $x$ as a
    function of $x$ for two different frequencies $\omega = 2$ and $\omega=6$ at times $t=234$ and $t=900$, respectively, calculated
    numerically with Newton method. Full lines are theoretical predictions for delocalized (diffusive) and localized modes given
    by Eqs.~(\ref{eq:Exdiff}, \ref{eq:Exloc}) with $R^2\approx 166$ and $R^2\approx 22$ correspondingly.}
    \label{fig:Ex}
\end{figure}
The difference between delocalized and localized modes is clearly
seen if we examine the dependence $E(x,t)$ as a function of
coordinate $x$ at some moment $t$ for two different frequencies
below and above the localization threshold. These two points for
investigation are shown on Fig.~\ref{fig:fan}. Black diamond
corresponds to delocalized mode with frequency $\omega=2$ and has
coordinates $t=234$ and $R^2=166$. The distribution of energy
$E(x,t)$ over the sample calculated numerically at this moment is
shown by black diamonds on Fig.~\ref{fig:Ex}. The data are
perfectly fitted by solid line drawn according to the solution of
diffusion equation in $1d$ case
\begin{equation}
    E(x,t) = \frac{E_{\rm tot}}{\sqrt{2\pi R^2}} \exp\left(-\frac{x^2}{2R^2}\right),
    \label{eq:Exdiff}
\end{equation}
with value of $R^2=166$.

Black triangle on Fig.~\ref{fig:fan} corresponds to localized mode
with frequency $\omega=6$ and has coordinates $t=900$ and
$R^2=22$. The distribution of energy $E(x,t)$ over the sample
calculated numerically at this moment is shown by black triangles
on Fig.~\ref{fig:Ex}. This distribution is drastically different
from the previous case. For localized modes we expect the usual
exponential decay
\begin{equation}
    E(x,t) = \frac{E_{\rm tot}}{\sqrt{2}R}\exp\left(-\frac{\sqrt{2}|x|}{R}\right).
    \label{eq:Exloc}
\end{equation}

The fit of the numerical data with this function and $R^2=22$ is
shown on Fig.~\ref{fig:Ex}. The fit is perfect except for the
central point at $x=0$ which lies noticeably above prediction of
Eq.~(\ref{eq:Exloc}). The coefficients in Eqs.~(\ref{eq:Exdiff},
\ref{eq:Exloc}) were taken to satisfy the obvious rules
\begin{equation}
\int\limits_{-\infty}^{\infty}E(x,t)dx=E_{\rm tot}, \quad
\frac{1}{E_{\rm tot}}\int\limits_{-\infty}^{\infty}x^2 E(x,t) dx =
R^2. \label{cf45}
\end{equation}

To find the diffusivity $D(\omega)$ for  $\mu\ne 0$, the method of
numerical solution of Newton's equations is not appropriate,
because in this case we have phonons in the lattice with long mean
free paths. Correspondingly samples with much bigger sizes are
necessary to use this approach. Therefore for $\mu\ne 0$ we used a
second approach. In this approach, the diffusivity $D(\omega_i)$
at eigenfrequency $\omega_i$  was calculated by means of the
formula of Edwards and Thouless~\cite{thouless}
\begin{equation}
    D(\omega_i) \simeq L^2 |\Delta\omega_i|
    \label{eq:ET}
\end{equation}
where $L$ is the length of the sample and $\Delta\omega_i$ is
sensitivity of the eigenfrequency $\omega_i$ to a twist of
boundary conditions. More precisely, we defined the diffusivity as
follows:
\begin{equation}
    D(\omega) = c\lim_{\varphi \to 0}\frac{L^2}{\varphi^2}\langle|\Delta\omega(\omega)|\rangle
     \label{eq:ETp}
\end{equation}
where $\varphi$ is the angle of twisting, and $c$ is some constant
of the order of unity. It will be determined from comparison with
the Newton method.  The averaging in Eq.~\ref{eq:ETp} is performed
over frequencies $\omega$ in the small interval
$(\omega-\delta\omega,\omega+\delta\omega)$ with
$\delta\omega=0.25$ and/or over  several thousands realizations.

\begin{figure}[h!]
     \includegraphics[scale=0.4]{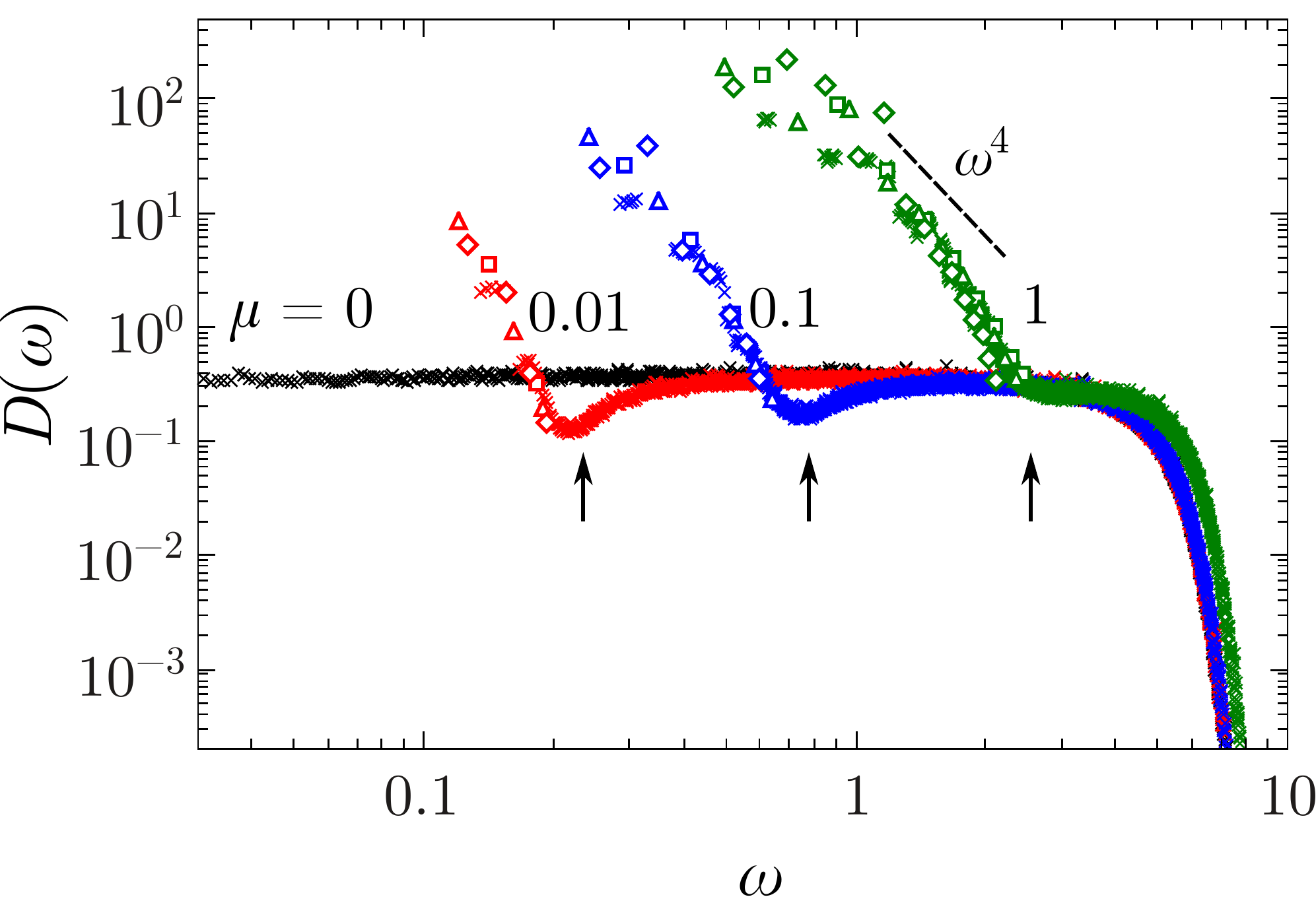}
     \caption{The diffusivity $D(\omega)$ for various $\mu$ (0, 0.01, 0.1, 1) for sample with $N=14^3$ (crosses).
     The diffusivity was calculated using formula of Edwards and Thouless (\ref{eq:ETp}) with $c=1$ and averaged over two
     thousand realizations. The arrows indicate frequencies $\omega_{\rm max}$ in the DOS $g(\omega)$ for corresponding values of $\mu$.
     Open symbols correspond to phonon diffusivity (\ref{9ngh}) below the Ioffe-Regel crossover frequency $\omega_{\rm IR}$.}
     \label{fig:Dw_mu}
\end{figure}
The symmetric real matrix $M$ was defined as usual (\ref{rt56e})
with periodic boundary conditions. The twisting of the matrix $M$
by angle $\varphi$ gives a new Hermitian matrix $M'$ obtained as
follows. For bonds between the left ($l$) and the right ($r$)
boundaries of our cubic sample
\begin{equation}
M'_{lr}=M_{lr}\exp(i\varphi),\quad M'_{rl}=M_{rl}\exp(-i\varphi) .
\end{equation}
For all other bonds $M'_{jk}=M_{jk}$. So $\Delta\omega_i$ is the
difference between $i$-th eigenfrequencies of matrices $M$ and
$M'$
\begin{equation}
    \Delta\omega_i = \omega_i - \omega'_i.
\end{equation}
Twisting of boundary conditions was performed for $x$ direction
only. For others two directions the periodic boundary conditions
were used.

For $\mu=0$ the results for $D(\omega)$ are shown on
Fig.~\ref{fig:Dw} for three different cubic samples (full lines).
We compared these results with numerical solution of Newton
equations for $\mu=0$ (black dots) and get for the constant $c
\approx 1$. Then we used this $c$ value for  $\mu\ne 0$. The
results are shown on Fig.~\ref{fig:Dw_mu}. For $\mu\ne 0$ we see
clearly  two different frequency regions in the function
$D(\omega)$.

At low frequencies, diffusivity increases with decreasing of
$\omega$. This range corresponds to the phonons. Indeed, the
diffusivity of phonons $D(\omega)$ can be calculated as follows
\begin{equation}
D(\omega) = \frac{1}{3}l(\omega)v_g(\omega) . \label{9ngh}
\end{equation}
Open symbols on Fig.~\ref{fig:Dw_mu} show contribution calculated
from this equation (just below Ioffe-Regel threshold). We see a
good agreement with Edwards and Thouless formula. After a deep
minimum at frequency $\omega\approx \omega_{\rm max}$ the
diffusivity $D(\omega)$ saturates at a constant level (independent
of $\mu$) coinciding with $D(\omega)$ for $\mu=0$. The diffusivity
in this range corresponds to diffusons. Similar behavior of
$D(\omega)$ was found recently in jammed systems~\cite{jammed1,
jammed2}. The deep minimum in the diffusivity at
$\omega\approx\omega_{\rm max}$ corresponds to strong scattering
of phonons by the quasilocal vibrations near the sharp peaks in
the DOS $g(\omega)$ (see Fig.~\ref{fig:gwmu}).

\section{Scaling relations}
\label{scaling}

Finally, the concept of diffusons allows us to establish useful
scaling relations between observable values and important
parameters of our model. One parameter is $\mu$. It has a
dimensionality of frequency squared. The second important
parameter of the model is the variance of non-diagonal elements
$A_{ij}$ of the random matrix $A$ which we hitherto considered to
be equal to unity
\begin{equation}
\left<A_{ij}^2\right>=V^2. \label{s6cv0}
\end{equation}
The parameter $V$ has dimension of frequency and assigns the scale
of typical frequencies in the system. In particular, the
normalized density of states $g(\omega)$ for $\mu=0$ shown on
Fig~\ref{fig:Dos1} has the following scaling relation
\begin{equation}
g(\omega)\simeq 1/V.
\end{equation}

Since for $\omega$ below $\omega_{\rm IR}$ in our disordered
lattice we have phonons  with $\omega=vq$ (here $v$ is sound
velocity) and above $\omega_{\rm IR}$ we have diffusons with
$\omega=Dq^2$ (here $D=D_{\rm rw}$) we can write at the
Ioffe-Regel threshold the order of the magnitude estimates
\begin{equation}
\omega_{\rm IR}\simeq vq_{\rm IR}, \quad \omega_{\rm IR}\simeq
Dq^2_{\rm IR}. \label{28bn}
\end{equation}
From these equations it follows that
\begin{equation}
v^2\simeq D\omega_{\rm IR}, \quad q_{\rm IR}^{-1} \simeq D/v.
\label{2ghb}
\end{equation}
Since, according to Eq.~(\ref{s7cvbf}) $v=\sqrt{E}$ (the units of
mass and length we put equal to unity, $m=a_0=1$), we find for the
Young modulus a useful relation
\begin{equation}
E\simeq D\omega_{\rm IR}\simeq D\sqrt{\mu} . \label{8c5g}
\end{equation}
Because, as we have shown in Section~\ref{diff}, the diffusion
coefficient $D$ is independent of $\mu$, the Young modulus has the
same $\mu$ dependence as $\omega_{\rm IR}\simeq\sqrt{\mu}$ (see
inset on Fig.~\ref{fig:gwmu}). It is in a full agreement with
Fig.~\ref{fig:Emu} for $\mu\ll 1$.

From the dimensionality considerations (since $V$ has a dimension
of frequency) we have for the diffusivity $D$ the following
estimate
\begin{equation}
D\simeq V .
\end{equation}
It is quite natural since the typical diffusion jump length is of
the order of lattice constant $a_0=1$ and typical jump frequency
is of the order of typical frequency in the system $V$. Therefore,
the diffusivity $D\simeq Va_0^2$. Taking this into account, for
the Young modulus and sound velocity at small frequencies we have
for $V\gg \sqrt{\mu}$
\begin{equation}
E\simeq V\sqrt{\mu} ,\quad v=\sqrt{E}\simeq (\mu V^2)^{1/4}
\end{equation}
i.e. the Young modulus is proportional to the characteristic
frequency in the system $V$. The correlation length (\ref{3b6g})
\begin{equation}
\lambda_{\rm IR} \simeq l(\omega_{\rm IR}) \simeq q^{-1}_{\rm IR}
\simeq \sqrt{D/\omega_{\rm IR}} \simeq D/v \simeq
\left(V^2/\mu\right)^{1/4} . \label{qkl9}
\end{equation}

Though our paper is not aimed at jamming transition and we
consider completely different model, it is interesting to note
that these scaling relations are identical to those found in
jamming transition~\cite{jammed2}. Authors~\cite{jammed2} study a
model of amorphous packing of frictionless spheres interacting via
the repulsive pair potential
\begin{equation}
U(r_{ij})\propto(1-r_{ij}/\sigma_{ij})^\alpha \quad \mbox{if}
\quad r_{ij}<\sigma_{ij} , \nonumber \label{99ff}
\end{equation}
\begin{equation}
U(r_{ij})=0 \quad \mbox{if} \quad r_{ij}>\sigma_{ij}, \label{a8cv}
\end{equation}
where the distance between the centers of particles $i$ and $j$ is
denoted by $r_{ij}$ and the sum of their radii by $\sigma_{ij}$.
This model system, irrespective of the value of $\alpha$, exhibits
a jamming/unjamming transition at $T=0$ at a packing fraction
$\phi=\phi_c$ at which the particles are just touching each other
and there is no overlap~\cite{hern}. At densities lower than
$\phi_c$ particles are free to rearrange while above $\phi_c$ at
$\Delta\phi \equiv \phi - \phi_c$, the system behaves as a weakly
connected amorphous solid with an average coordination number that
scales as a power law with an exponent
\begin{equation}
\Delta z\equiv z-z_c\sim \Delta\phi^{1/2}
\end{equation}
where $z_c=2d$, with $d$ being the space dimension.

It was found that different quantities exhibit scaling behavior
near the jamming point. According to~\cite{jammed2} the
Ioffe-Regel crossover frequency $\omega^*$ and the shear modulus
$G$ behave as (we use below the notation of the
paper~\cite{jammed2})
\begin{equation}
\omega^* \sim \Delta\phi^{(\alpha-1)/2}, \quad G\sim
\Delta\phi^{(2\alpha-3)/2}. \label{a6mn}
\end{equation}
The transverse sound velocity $v_t$ and the diffusivity in the
plateau region $d_0$ scale
\begin{equation}
v_t\sim\Delta\phi^{(2\alpha-3)/4} , \quad d_0\sim
\Delta\phi^{(\alpha-2)/2}. \label{abc5}
\end{equation}
The applied pressure $p$ and the plateau in the density of states
$D_0$ depend on the packing fraction as follows~\cite{hern}
\begin{equation}
p\sim \Delta\phi^{\alpha-1}, \quad D_0\sim
\Delta\phi^{(2-\alpha)/2}. \label{a7xc}
\end{equation}

Thus if we put
\begin{equation}
\mu\sim\Delta\phi^{\alpha-1}, \quad V\sim\Delta\phi^{(\alpha-2)/2}
, \label{s7vb}
\end{equation}
then the crossover frequency $\omega_{\rm IR}$, the Young modulus
$E$, sound velocity $v$, the diffusivity at the plateau $D$, and
the density of states $g(\omega)$ in our model have the same
scaling as the crossover frequency $\omega^*$, the shear modulus
$G$, transverse sound velocity $v_t$, the diffusivity in the
plateau $d_0$, and the density of states $D_0$ in the jamming
transition model respectively. In particular, the parameters $\mu$
and $V$ in our model are equivalent to pressure $p$ and inverse
density of states $1/D_0$ in the jamming transition model
correspondingly.

In the paper we mainly considered a case of strong disorder, $\mu
\ll V^2$. Taking into account Eq.~(\ref{s7vb}) we find that the
small parameter of our model
\begin{equation}
\mu/V^2 \sim \Delta\phi \label{6x7}
\end{equation}
coincides with the small parameter $\Delta\phi$ in the jamming
transition model. The mean free path at the crossover as follows
from (\ref{qkl9}) and (\ref{s7vb}) is given by
\begin{equation}
l(\omega_{\rm IR})\sim \Delta\phi^{-1/4} ,
\end{equation}
what also coincides with~\cite{jammed2}. It would be very
interesting to investigate physical reasons for this striking
``mapping'' of two models to each other in more details in a future
work.

\section{Discussion}
\label{disc}

We have developed a stable random matrix approach to describe vibrations
in strongly disordered systems, which have properties similar to
what one observes in granular matter at the jamming transition
point, in jammed systems and, finally, in real glasses. This
approach has one important advantage in comparison to other
models. It describes mechanical systems which are always stable
independently of the degree of disorder. Previous random matrix
models~\cite{schirm, taraskin3, grigera} suffer  from an inherent
mechanical instability that occurs at some critical amount of
disorder. As a result they are limited by consideration of
"relatively weak" or "moderate" disorder.

We use scalar model and take the dynamical matrix in the form
$M=AA^T+\mu M_0$. Here $A$ is a random matrix $N\times N$ built on
a simple cubic lattice with $N$ particles and interaction between
nearest neighbors only. The only non zero non-diagonal matrix
elements $A_{ij}$ between the nearest neighbors are taken as
independent random numbers from Gaussian distribution with zero
mean $\left<A_{ij}\right>=0$ and unit variance
$\left<A^2_{ij}\right>=V^2=1$. The variance controls the degree of
disorder in the lattice.  To ensure the important property
(\ref{s6cv1}) the diagonal elements are calculated as a minus sum
of non-diagonal elements $A_{ii}= -\sum_{j\ne i} A_{ji}$. $M_0$ is
a  crystalline dynamical matrix with unit springs between the
nearest neighbors. As a result each particle in this lattice is
connected by random elastic springs with 24 surrounding particles.
Since matrix $AA^T$ is always positive definite, such form of the
dynamical matrix guarantees the mechanical stability of the system
for any positive value of $\mu$.

If the first term $AA^T$ is responsible for the disorder in the
system, the second term $\mu M_0$ describes the ordered part of
the Hamiltonian. The parameter $\mu$ controls the relative
amplitude of this part and the rigidity of the lattice. It  can
vary in the interval $0\leqslant \mu < \infty$, changing the
rigidity and relative amount of disorder. In this paper we have
mainly considered the case of strong and moderate disorder when
$0\leqslant \mu\lesssim V^2$ and fluctuating part of the
dynamical matrix is bigger then the ordered part.  In this case
the Young modulus of the lattice $E\propto V\sqrt{\mu}$. The
parameter $\mu$ plays the same role as pressure in jammed systems.

We have found that the delocalized vibrational excitations in this
disordered lattice are of two types. At low frequencies below the
Ioffe-Regel crossover, $\omega < \omega_{\rm IR}$, they are the
usual phonons (plane waves) which can be characterized by
frequency $\omega$ and wave vector $\bf q$. However, with
increasing of $\omega$, due to the disorder-induced scattering,
the phonon line width $\Delta\omega$ increases rapidly as
$\Delta\omega\propto\omega^{4}$ and at some frequency
$\omega\approx\omega_{\rm IR}$ the phonon mean free path $l$
becomes of the order of the wave length $\lambda$. Though this
crossover is not sharp and has no critical behavior at
$\omega=\omega_{\rm IR}$, the structure of the eigenmodes at
higher frequencies quite soon become very different from the plane
waves.

As a result, at higher frequencies the original notion of phonons
is lost and delocalized vibrational modes have a diffusive nature.
They are similar to {\em diffusons} introduced by Allen and
Feldman, et al.~\cite{Nature5}. The diffusons again can be
characterized by frequency $\omega$, but have no well defined wave
vector $\bf q$. Above $\omega\approx\omega_{\rm IR}$ the structure
factor of particle displacements $S({\bf q}, \omega)$  becomes
very similar to the structure factor $S_{\rm rw}({\bf q}, \omega)$
of a random walk on the lattice.  The former has a broad maximum
as a function of $q$ at $q=\sqrt{\omega/D_u}$, where $D_u\simeq V$
is a diffusion coefficient of the particle displacements.

The displacement structure factor $S(q,t)$ in the diffuson range,
for small $q\ll 1/a_0$,  decays as following,
$S(q,t)\propto\exp(-D_uq^2t)$. As a result the vibrational line
width $\Gamma(q)=D_u q^2$. Such quadratic dependence of
$\Gamma(q)$ was found in many glasses in the experiments on
inelastic x-ray scattering, see for example~\cite{sette,
ruoccosette} and references therein. It was also found in
molecular dynamic simulation of amorphous silicon~\cite{christie}.
However in these and other papers this line width was attributed
to phonons without discussion of its physical origin. We guess
that the observed $q^2$ dependence of $\Gamma(q)$ has nothing to
do with phonons and is in fact related to diffusons.  However, a
more detailed investigation is necessary for a definite
conclusion.

The crossover between phonons and diffusons takes place at the
Ioffe-Regel crossover frequency $\omega_{\rm IR}$ which is close
to the position of the boson peak. Since for phonons
$\Delta\omega\propto\omega^4$ and for diffusons
$\Gamma(q)=D_uq^2$, there should exist a crossover from $\omega^4$
to $q^2$ dependence of the line width. Such a crossover was indeed
found recently in inelastic x-ray scattering in lithium diborate
glass~\cite{laermans3}, densified vitreous
silica~\cite{ruffle2003}, vitreous silica~\cite{valentina1,
valentina2, valentina3}, glassy sorbitol~\cite{valentina4} and
glycerol glass~\cite{valentina5}. The crossover frequency was
found to be close to the BP position.

As a result, if our guess is true, we can calculate the diffusion
coefficient of particle displacements, $D_u=\Gamma(q)/q^2$, from
the experimental line width $\Gamma(q)$ in the range, where it is
proportional to $q^2$. Taking into account that $D_u\approx
a_0^2/\tau$ where $a_0$ is the lattice constant and $\tau$ is an
average time for a jump, we come to the order of the value
estimate $D_u\approx 1$\,mm$^2$/sec for $a_0\approx 2$\,\AA\, and
$\tau\approx 0.4\times 10^{-13}$\,sec. Let us compare this value
with experimental data.

In the paper~\cite{valentina2} it was found that in vitreous
silica $\hbar\Gamma/(\hbar\omega)^2=0.07$\,meV$^{-1}$ for $q\ge
2$\, nm$^{-1}$. Taking the sound velocity $v_L=5250$\,
m\,sec$^{-1}$ for $q=2$\, nm$^{-1}$ we get for diffusion
coefficient $D_u=1.3$\,mm$^2$/sec. Let us compare this value with
the diffusivity of energy $D(\omega)$ for small $\omega$ in the
same glass. We expect that both coefficients should be of the same
order of magnitude. The diffusivity of energy $D(\omega)$ in
vitreous silica was calculated in the paper~\cite{sim}. It was
obtained that $D(0)=1.4$\,mm$^2$/sec. A close estimate
$D(0)=1.1$\,mm$^2$/sec was given in~\cite{similar}. As one can see
the agreement between $D_u$ and $D(0)$ is unexpectedly  good. In
glycerol glass~\cite{ruocco} we found the diffusivity about factor
of two  smaller, $D_u=0.46$\,mm$^2$/sec. For amorphous silicon
from molecular dynamic calculations~\cite{christie} we get
$D_{ul}=3.2$\,mm$^2$/sec for longitudinal vibrations, and
$D_{ut}=1.2$\,mm$^2$/sec for transverse vibrations. For the
diffusivity of energy we have in this glass the
estimate~\cite{Nature5} $D(0)=0.6$\,mm$^2$/sec.

Since $\omega_{\rm IR}\propto\sqrt{\mu}$ (and independent
of the strength of disorder $V$), we can vary the
Ioffe-Regel crossover frequency and, therefore, the relative
number of phonons $N_{\rm ph}$ in the system, changing the
parameter $\mu$.  It is zero when $\mu=0$ and there are no phonons
in the lattice. In this case all delocalized vibrations are
diffusons.  If $0 <\mu\ll 1$ we have phonons, but their relative
number is small. One can show that in this case $N_{\rm
ph}\propto\mu^{3/4}$. In the opposite case, $\mu\gg 1$,  the
disorder is relatively small and nearly all vibrations in the
lattice are well defined plane waves, i.e. phonons.

In amorphous silicon the relative number of phonons (plane waves)
was estimated to be only 4\% from all of the vibrational modes in
the system~\cite{Nature5}. The estimates show that in our model we
have such a small amount of propagating modes, as in a-Si, for
$\mu\approx 0.1$. In the silica glass we can estimate the relative
number of phonons from the data~\cite{taraskin}. Taking into
account that Ioffe-Regel crossover frequency in amorphous silica
was estimated to be~\cite{taraskin} $\nu_{\rm IR}=1$\,THz, and
integrating density of states~\cite{taraskin} up to this frequency
we come to the relative number $N_{\rm ph}=0.002\pm 0.0005$.  As a
result in the typical glass such as amorphous silica only
$0.2$\,\% of all modes are phonons. As follows from
Table~\ref{tab1} it corresponds to very small values of $\mu
<0.01$. It means that small amount of phonons in disordered
systems is a signature of strong disorder.

Usually the phenomenon of diffusion takes place for conserved
quantities. In our system we have two integrals of motion. They
are the momentum and the energy of the lattice. Therefore, first
of all, one has to discriminate  the diffusion of particle
momentums (or particle displacements) from the diffusion of
energy.  Conservation of displacement is related to conservation
of the center of inertia in the system. As a result, the diffusion
of particle displacements has the same diffusion coefficient as
the diffusion of particle momentums.

The diffusion coefficient of displacements/momentums $D_{u/v}$ is
hidden in the displacement structure factor $S({\bf q}, \omega)$
(\ref{45gto}). Comparing this structure factor with the structure
factor of  the random walk on the lattice, we found that for the
case of $\mu=0$ the diffusion coefficient $D_{u/v}=D_{\rm
rw}=0.7$. We can check that it  is indeed the diffusion
coefficient of particle displacements/momentums in a similar way
we used for finding the diffusivity  of energy $D(\omega)$ in
Section~\ref{ub55p}.

Let us consider a cubic random lattice $L\times L\times L$ with
$\mu=0$ and unit masses $m_i=1$ with periodic boundary conditions.
At initial moment $t=0$ let us displace all particles in a thin
layer around the central layer (with coordinate $x=0$) according
to Gaussian distribution
\begin{equation}
u(x,0)=u_0 e^{-x^2/2x_0^2}. \label{ts45n}
\end{equation}
Here the thickness of the layer $x_0$ should be small enough in
comparison to the sample size $L$, i.e.  $x_0\ll L/2$. Initial
velocities ${\dot u}(0)$ of all the particles are equal to zero.

After initial displacements in the thin central layer, the
particle displacements will diffuse to the left and to the right
ends of the sample. Solving numerically the Newton equations, we
find the average squared distance to the displacement diffusion
front, similar to Eq.~(\ref{eq:NR2})
\begin{equation}
R^2_u(t)=\frac{1}{u_{\rm tot}} \sum\limits_i x_i^2 u_i(t), \quad
u_{\rm tot}=\sum\limits_i u_i(t). \label{et55}
\end{equation}
Since the center of inertia does not move, the total displacement
of all particles $u_{\rm tot}$ is independent of time and equal to
the total displacement at $t=0$.

From the slope of $R^2_u(t)$ we can calculate the diffusion
coefficient of the displacements $D_u$ as follows
\begin{equation}
R_u^2(t)=2D_u t \label{x5as}
\end{equation}
similar to Eq.~(\ref{eq:ND}).

In the same way we can calculate the diffusion of momentum. For
that at the moment $t=0$ initial displacements of all the
particles we put equal to zero. However  initial velocities
$v={\dot u}(0)$ in the thin central layer we take distributed
similar to Eq.~(\ref{ts45n})
\begin{equation}
v(x)=v_0 e^{-x^2/2x_0^2}. \label{ts46n}
\end{equation}
Then, as in the previous case, solving numerically the Newton
equations we find
\begin{equation}
R^2_v(t)=\frac{1}{v_{\rm tot}} \sum\limits_i x_i^2 v_i(t), \quad
v_{\rm tot}=\sum\limits_i v_i(t). \label{et56}
\end{equation}
Since the total momentum is conserved, $v_{\rm tot}$ is also
independent of time and equal to its initial value at $t=0$. From
the slope of $R^2_v(t)$ we can calculate the diffusion coefficient
of the momentum $D_v$ using one dimensional equation
\begin{equation}
R_v^2(t)=2D_v t \label{x5asd}
\end{equation}
similar to Eq.~(\ref{x5as}).

In both cases we have obtained for diffusion coefficients $D_u$
and  $D_v$ the same value as was derived from the structure
factor, $D_u\approx D_v\approx D_{\rm rw}=0.7$. It confirms our
statement that the displacement structure factor $S({\bf q},
\omega)$ gives us the information about diffusion of particle
displacements (or momentums). The diffusion of momentum is usually
related to viscosity $\eta$ of the medium. Therefore in the case
of $\mu=0$ our lattice  has no rigidity but has a finite value of
viscosity.

In disordered lattices the diffusion of energy is different from
the diffusion of particle displacements (momentums). In the
harmonic approximation the eigenmodes with different frequencies
do not interact with each other. Therefore the energy  cannot be
transferred from one eigenmode to other eigenmodes. It means that
energy of every eigenmode $E(\omega_i)$ is conserved (with time).
The total energy $E_{\rm tot}$ is just a sum of these eigenmode
contributions
\begin{equation}
E_{\rm tot}=\sum\limits_iE(\omega_i). \label{x5a9}
\end{equation}
As a result, instead of one integral of motion (the total energy
$E_{\rm tot}$), in a scalar harmonic system with $N$ particles we
have $N$ integrals of motion $E(\omega_i)$. And for each frequency
$\omega_i$ we have its own unique energy diffusivity
$D(\omega_i)$. At this point our model decidedly confirms the
physical picture suggested in papers~\cite{Nature1, Nature2,
Nature3, Nature4, Nature5} for amorphous silicon. We believe that
it can be applied to some other glasses as well.

Usually this diffusivity is hidden in a displacement/momentum
structure factor of the $4$-th order. However, we calculated the
diffusivity of energy $D(\omega)$ in a different way using two
different approaches as it was discussed in Section~\ref{ub55p}.
The first approach is based on the direct solution of Newton
equations. In the second approach we calculated the diffusivity
using Edwards and Thouless formula~\cite{thouless}. Both
approaches give the same result.

In the first approach we used a short external force pulse $\Delta
t$ exciting vibrations in a small space region of the lattice and
in a small frequency interval $\Delta\omega\approx 1/\Delta t$
near frequency $\omega$.  Then on a time scale $t\gg \Delta t$ the
energy  diffused through the lattice.  Using Newton equations of
motion we calculated this diffusion directly. It was supposed that
the interval $\Delta\omega$ is much bigger than the interlevel
spacing $\delta\omega$ and therefore the former consists of many
eigenmodes. In the thermodynamic limit $\delta\omega\propto 1/N\to
0$ if  $N\to\infty$.  Therefore in an infinite system we can take
the interval $\Delta\omega$ arbitrary small. The energy diffusion
coefficient $D(\omega)$ in this case is a function of frequency
$\omega$. Approaching the localization threshold $\omega_{\rm
loc}$ the diffusivity $D(\omega)$ should go to zero.

We applied this method for $\mu=0$, when there are no phonons in
the lattice. In this case we obtained for diffusivity at zero
frequency $D(0)\approx 0.4$, i.e. the value about factor of two
smaller than for diffusivity of displacements, $D_u$. However this
approach is rather difficult to implement  for computer
simulations in the case when $\mu\ne 0$. In this case we have
phonons in the lattice with long mean free paths. And samples with
much bigger sizes are necessary.

Therefore, to calculate the diffusivity $D(\omega)$ for arbitrary
value of $\mu$ (including the case of $\mu=0$), we used another
approach.  In this approach Edwards and Thouless
formula~\cite{thouless}, $D(\omega_i) = c L^2 |\Delta\omega_i|$,
was used. It relates the diffusivity $D(\omega_i)$ with shift of
the eigenfrequencies $\Delta\omega_i$ due to change of the
boundary conditions in one direction. The proportionality
coefficient $c$ we found from the comparison with the Newton
method for $\mu=0$. In this case both methods result in the same
frequency dependence of $D(\omega)$.

The diffusivity of vibrational modes $D(\omega)$ in disordered
lattices is a very important quantity. It determines the thermal
conductivity~\cite{Nature2,jammed1,jammed2}
\begin{equation}
\varkappa(T) \propto \int\limits_0^\infty d\omega
g(\omega)D(\omega)C(\omega, T). \label{thcond1}
\end{equation}
Here $g(\omega)$ is density of states and $C(\omega,T)$ is
specific heat of harmonic oscillator
\begin{equation}
C(\omega,T)=\left(\frac{\hbar\omega}{T}
\right)^2\frac{e^{\hbar\omega/T}}{\left(e^{\hbar\omega/T}-1
\right)^2}. \label{specificheat1}
\end{equation}
Localized modes have $D(\omega_i)=0$ and make no contribution to
$\varkappa(T)$.

If functions $g(\omega)$ and $D(\omega)$ are approximately
constant in some frequency interval (the case that we have, for
example, in our picture for $\omega>\omega_{\rm IR}$), then we
find from Eq.~\ref{thcond1} that approximately
$\varkappa(T)\propto T$ in the corresponding temperature
range~\cite{jammed1}. It explains a quasi-linear temperature
dependence of the  thermal conductivity above the plateau observed
in glasses~\cite{CahillPohl}. With increasing frequency the
functions $g(\omega)$ and $D(\omega)$ finally drop to zero and
thermal conductivity saturates at some constant level independent
of temperature. Thus the conception of diffusons gives clear
explanation for the temperature dependence of the thermal
conductivity of glasses and other disordered systems.

Summarizing, using a stable random matrix approach we have
presented a consequent theory of vibrational properties in
strongly disordered systems. In these systems a relative amount of
phonons is small and  almost all delocalized vibrations are
diffusons. The diffusons play an important role and are
responsible for the transport properties of glasses at higher
temperatures. Presumably they are also accounted for the
mysterious $q^2$ dependence of the vibrational line width
$\Gamma(q)$  observed in many experiments on inelastic x-ray
scattering in glasses. Therefore we think that it is necessary to
take them into account in interpretation of experimental data.

\section{Acknowledgments}

We are very grateful to V.~L.~Gurevich  and Anne Tanguy for many
stimulating discussions and gratefully acknowledge interesting
discussions with B. Ruffl\'e and E.~Courtens as well. One of the
authors (DAP) thanks the University Lyon 1 for  hospitality. This
work was supported by St. Petersburg Government (diploma project
no. 2.4/29-06/143C), Dynasty Foundation, RF President Grant
``Leading Scientific Schools'' NSh-5442.2012.2 and Russian
Ministry of Education and Science (contract N~14.740.11.0892).

\section{Appendices}

\subsection{Lattices with cut out bonds}
\label{cutlat}

\begin{figure}[!h]
    \includegraphics[scale=0.4,keepaspectratio]{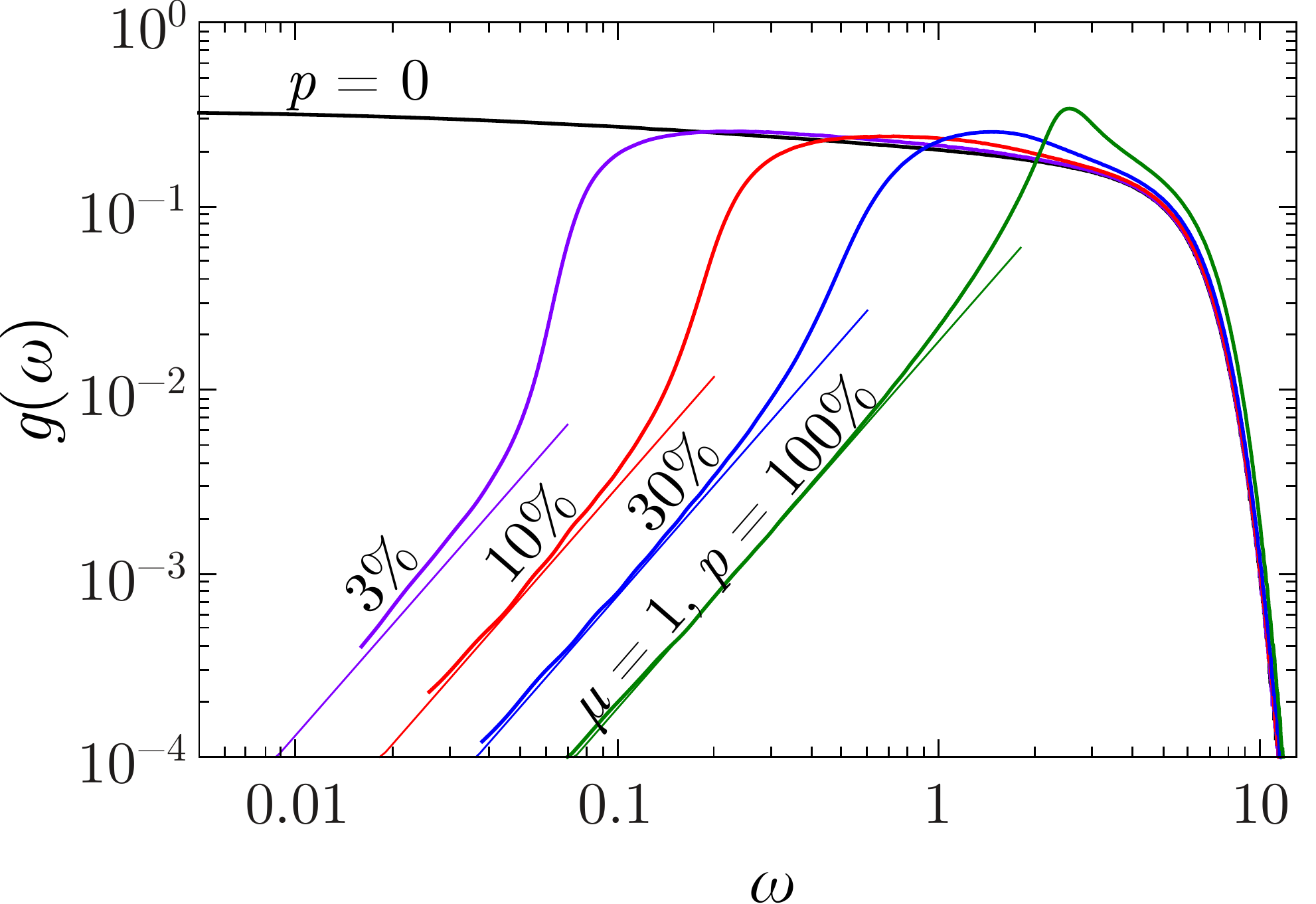}
    \caption{The normalized DOS $g(\omega)$ for dynamical matrix $M=AA^T+\mu M_0$ with $\mu=1$ and different percentage $100\%-p$ of
    cut out springs calculated with precise numerical KPM  solution for cubic lattice with $N=200^3$ (full lines). Straight lines
    are calculated according to Eq.~(\ref{st56}) with sound velocity $v=\sqrt{E}$. The Young modulus $E$ is calculated in the same
    way as in the Section~\ref{phonons}.}
    \label{fig:cut}
\end{figure}
Consider here the case when some part of springs $\mu$ are cut out
from the matrix $\mu M_0$ in dynamical matrix (\ref{rt56e}). The
value of parameter $\mu=1$ we will keep fixed. Let parameter $p$
gives the percentage of remaining springs.
The percolation threshold in the simple cubic lattice for
bond percolation problem
is at $p_c\approx 25\%$~\cite{percolation}.  If $p<p_c$, then
there is no infinite cluster of connected springs and therefore
matrix $\mu M_0$ with cut out springs itself has no acoustical phonon-like modes
at all. Nevertheless, the full dynamical matrix (\ref{rt56e})
still has well defined phonon modes with density of states
$\propto\omega^2$ for all positive values of $p$ even below the
percolation threshold. The normalized density of states
$g(\omega)$ for $\mu=1$ and different values of $p$ is shown on
Fig.~\ref{fig:cut}. The straight lines show the phonon
contribution to the DOS calculated from Eq.~(\ref{st56}) with
sound velocity given by Eq.~(\ref{s7cvbf}). The Young modulus $E$
was calculated numerically using Eq.~(\ref{d6vg}) for the lattice
with $N=10^6$ particles (one realisation) in the same way as it
was done in Section~\ref{phonons}. The details of these calculations
will be published elsewhere.

\subsection{Superposition of two random matrices}
\label{superposition}

Another (less obvious) possibility to get phonons is to add to the random dynamical matrix $AA^T$ a random matrix
$\beta BB^T$. Here $\beta$ is a parameter and the random matrix $B$ is build in the same way as random matrix $A$ but they are statistically independent from each other. Though both terms $AA^T$ and $\beta BB^T$ taken separately have zero rigidity (and do not have phonons) their superposition introduces a finite rigidity
$E$ to the system. The rigidity changes when we vary parameter $\beta$ as $E\propto\sqrt{\beta}$
and goes to zero when $\beta\to 0$. So the scaling relations
\begin{figure}[!h]
    \includegraphics[scale=0.4,keepaspectratio]{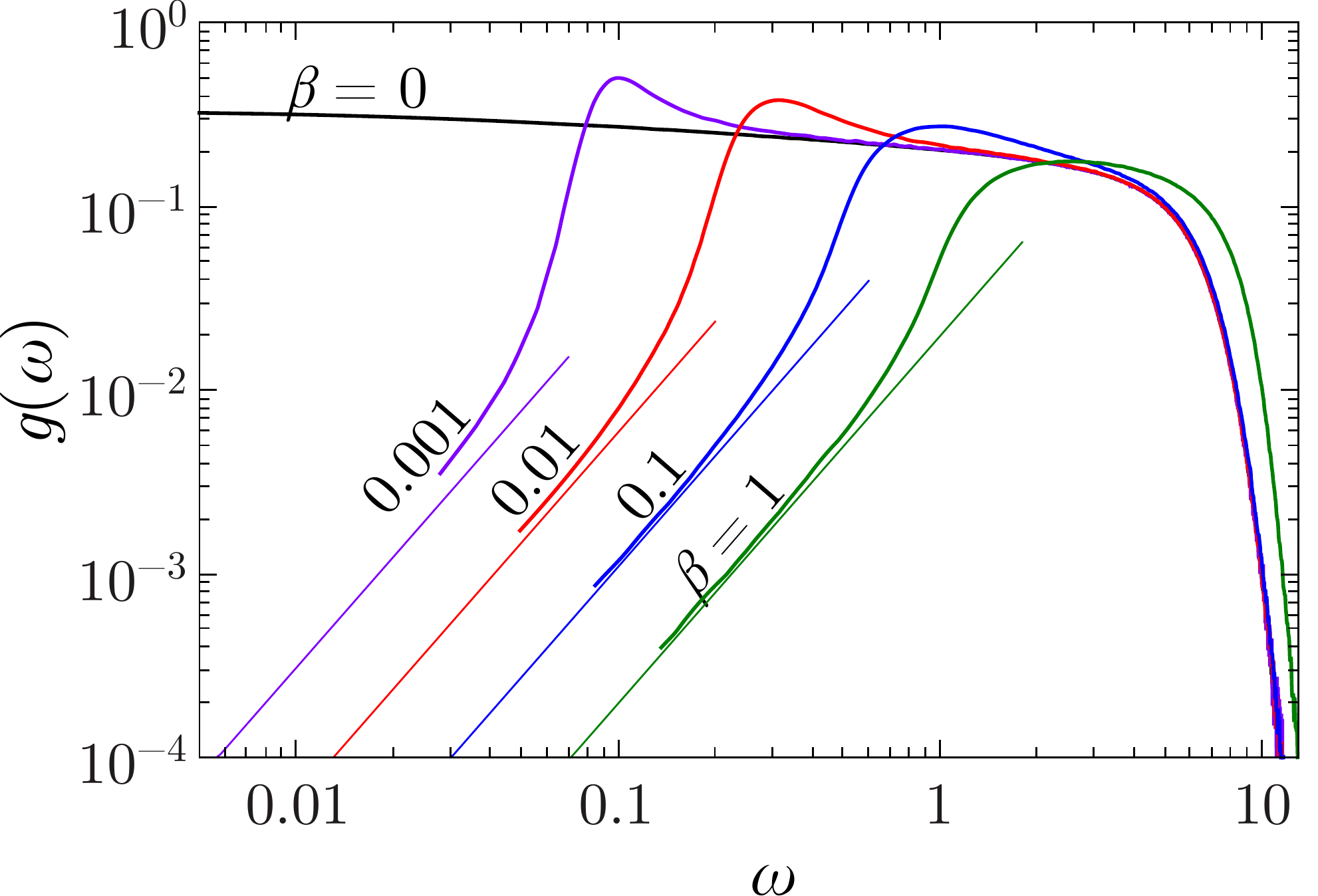}
    \caption{The normalized DOS $g(\omega)$ for dynamical matrix $M=AA^T+\beta BB^T$ with different $\beta$  calculated with precise numerical KPM  solution for simple  cubic lattice with $N=100^3$ (full lines). Straight lines are calculated according to Eq.~(\ref{st56}) with sound velocity $v=\sqrt{E}$. The Young modulus $E$ is calculated in the same way as in the Section~\ref{phonons}.}
    \label{fig:beta}
\end{figure}
in this case for $\beta\ll V^2$ are the
same as in Section~\ref{scaling} with replacement of $\mu$ by $\beta$.
The preliminary results obtained within this approach are shown
on Fig.~\ref{fig:beta}. Further details will be published elsewhere.

These two examples show clearly, that appearance of phonons in the system
is not related to the crystalline order in the term $\mu M_0$. The issue is more complicated.
We are going to discuss this problem in more details elsewhere.

\subsection{Displacement structure factor}
\label{DSF}

Let us consider the displacement structure factor given by
Eq.~(\ref{45gto})
\begin{equation}
S({\bf q}, \omega) =
\frac{2}{NT}\left|\sum\limits_{i=1}^Ne^{-i{\bf q}{\bf
r}_i}\int\limits_0^{T} u({\bf r}_i, t) e^{i\omega t}dt \right|^2 .
\label{45gto1}
\end{equation}
We will assume that initial velocities of all particles at $t=0$
are zero. Then the displacement of $i$-th particle $u({\bf r}_i,
t)$ as a function of time can be written in the form
\begin{equation}
    u({\bf r}_i, t) = \sum\limits_{j=1}^N a_j e_i(\omega_j) \cos(\omega_j t).
    \label{eq:ut}
\end{equation}
Here $e_i(\omega_j)$ --- is eigenvector of the dynamical matrix
$M$ corresponding to $i$-th particle and eigenfrequency
$\omega_j$. The eigenvectors satisfy equations
\begin{equation}
\sum\limits_{j=1}^N M_{ij}e_j(\omega_k)=\omega_k^2e_i(\omega_k).
\label{s66f}
\end{equation}
They form an orthogonal set~\cite{maradudin}, so that
\begin{equation}
\sum\limits_{j=1}^N e_{i}(\omega_j)e_{k}(\omega_j)=
\sum\limits_{j=1}^N e_j(\omega_i)e_j(\omega_k) = \delta_{ik}.
\label{z7bv}
\end{equation}
Using (\ref{z7bv}), one can write the coefficients $a_j$ in
(\ref{eq:ut}) in terms of the particle displacements for $t=0$
\begin{equation}
a_j=\sum\limits_{i=1}^N u({\bf r}_i, 0) e_i(\omega_j).
\label{s55cv}
\end{equation}
The initial displacements $u({\bf r}_i, 0)$ are independent
Gaussian random variables with zero mean and unit variance
\begin{equation}
\left<u({\bf r}_i, 0)\right>=0, \quad \left<u({\bf r}_i, 0)u({\bf
r}_j, 0)\right>=\delta_{ij} . \label{a6xc}
\end{equation}
Basing on this equation and making use of (\ref{s55cv}) and of
(\ref{z7bv}) one can prove that the coefficients $a_j$ are also
independent random Gaussian variables
\begin{equation}
\left<a_j\right>=0, \quad \left<a_i a_j\right>=\delta_{ij}.
\label{q1b6}
\end{equation}
Using this property, one can evaluate the average (\ref{45gto1})
as
\begin{equation}
\left<S({\bf q}, \omega)\right>=\frac{2}{NT}\sum\limits_{j=1}^N
\left|\sum\limits_{i=1}^N e_i(\omega_j) e^{-i{\bf q}{\bf r}_i}
\right|^2 \left|\int\limits_0^T \cos(\omega_j t)e^{i\omega
t}dt\right|^2 . \label{88vb}
\end{equation}
Having in mind that
\begin{equation}
    \lim_{T\to \infty}\frac{2}{T}\left|\int\limits_0^T\cos(\omega_jt) e^{i\omega t} dt \right|^2 = \pi \left(\delta(\omega-\omega_j)+\delta(\omega+\omega_j)\right)
\end{equation}
and taking only positive frequencies, we arrive to
\begin{equation}
    \left<S({\bf q}, \omega)\right> = \frac{\pi}{N}\sum\limits_{j=1}^N\left|\sum\limits_{i=1}^N
    e_i(\omega_j)e^{-i{\bf q}{\bf r}_i}\right|^2\delta(\omega-\omega_j).
    \label{eq:VSqw2}
\end{equation}



\begin{thebibliography}{99}
\bibitem{fluct} We have checked that for $\mu > 10^{-4}$ for such a big sample the fluctuations of Young modulus from sample to sample are small so we can use one realization only. It is different from the case $\mu=0$ where the relative fluctuations of the Young modulus are of the order of unity~\cite{our2}.
\bibitem{central} For big sample with $N=100\times 100\times 100=10^6$ particles it is sufficient to excite only one atomic layer $x=0$ with $100\times 100 = 10^4$ particles. Addition of two or more neighbor layers does not change the results. Increasing the width of the excited layer one should increase the length of the sample as well.
\end{thebibliography}
\end{document}